 \documentclass[final,5p,times,twocolumn]{elsarticle}

\usepackage{graphicx}
\usepackage{subfigure}
\usepackage{multirow}
\usepackage{wrapfig}
\usepackage{array}

\usepackage{tabularray}
\usepackage{tabularx}
\usepackage{rotating}
\usepackage{makecell}
\usepackage{booktabs}

\usepackage[utf8]{inputenc}
\usepackage{amsmath}
\usepackage{amsthm}
\usepackage{amsfonts}
\usepackage{epsfig}
\usepackage{psfrag}
\usepackage{pstricks}
\usepackage{algorithm}
\usepackage{stfloats}

\usepackage{cuted}

\usepackage{algorithmicx}
\usepackage{algpseudocode}  
\floatname{algorithm}{Algorithm}

\graphicspath{{./Figures/}}

\usepackage{amssymb}
\usepackage{bm}
\usepackage{hyperref}
\hypersetup{
            colorlinks=true,
            linkcolor=blue,
            anchorcolor=blue,
            citecolor=blue
            }

\biboptions{comma,sort&compress}

\usepackage{enumitem}
\setlist{nosep,topsep=-\parskip}

\usepackage{ulem}

\journal{Elsevier}

\begin{document}

\begin{frontmatter}

\title{A review of geometric modeling methods in microstructure design and manufacturing}

\author[]{Qiang Zou\corref{cor}}\ead{qiangzou@cad.zju.edu.cn}
\author[]{Guoyue Luo}

\cortext[cor]{Corresponding author.}
\address{State Key Laboratory of CAD$\&$CG, Zhejiang University, Hangzhou, 310058, China}

\begin{abstract}
Microstructures, characterized by intricate structures at the microscopic scale, hold the promise of important disruptions in the field of mechanical engineering due to the superior mechanical properties they offer. One fundamental technique of microstructure design and manufacturing is geometric modeling, which generates the 3D computer models required to run high-level procedures such as simulation, optimization, and process planning. There is, however, a lack of comprehensive discussions on this body of knowledge. The goal of this paper is to compile existing microstructure modeling methods and clarify the challenges, progress, and limitations of current research. It also concludes with future research directions that may improve and/or complement current methods, such as compressive and generative microstructure representations. By doing so, the paper sheds light on what has already been made possible for microstructure modeling, what developments can be expected in the near future, and which topics remain problematic.
\end{abstract}


\end{frontmatter}

\stripsep -18pt
\begin{strip}
\begin{minipage}{\textwidth}
     \setcounter{tocdepth}{2}
    \tableofcontents
\end{minipage}
\end{strip}

\clearpage
\newpage

\section{Introduction}
\label{sec:intro}
\begin{figure*}[b]
\centering
\includegraphics[width=0.9\textwidth]{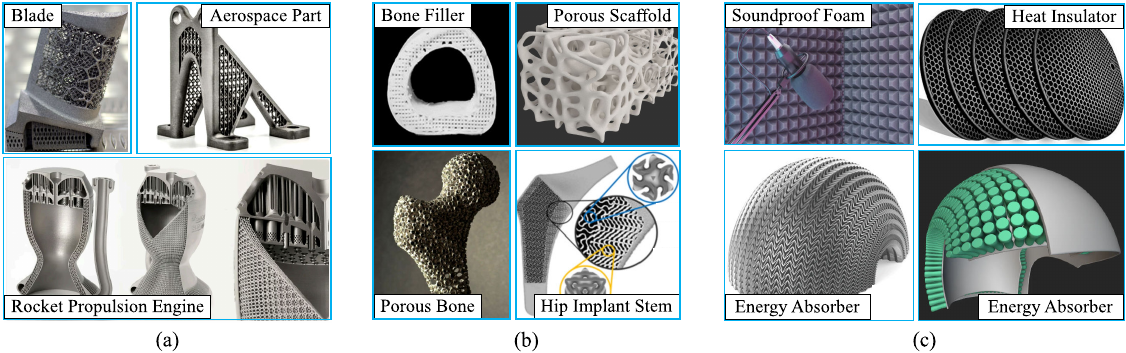}
\caption{Typical microstructure applications: (a) aerospace industry; (b) biomedical industry; and (c) applications in heat, noise canceling, or energy absorption.}
\label{fig:application}
\end{figure*}

Additive manufacturing (AM) constructs 3D objects by accumulating materials, layer by layer. In principle, this additive process can make parts with arbitrary shape complexity and material composition~\cite{2023_Singh_review_AM_Forward-Looking}. 
For this reason, AM has found extensive applications in industries such as aerospace, automotive,
and architecture~\cite{2020_AM-application-mass-production,2016_Thompson_revirw_AM_CIRP,2020_AM-application-building}. 
One particular area promoted by AM is the design and manufacturing of microstructures, which have intricate structures at the microscopic scale. Structures of this kind can provide superior mechanical properties such as lightweight, high strength, and multifunctional, which are advantageous in aerospace, biomedical, and energy applications~\cite{2017_Qin_lattice-mechanical-property,2021_Obadimu_lattice_compressieve-behavior}. Fig.~\ref{fig:application} illustrates examples of microstructure applications.

\begin{figure}[htbp]
\centering
\includegraphics[width=0.48\textwidth]{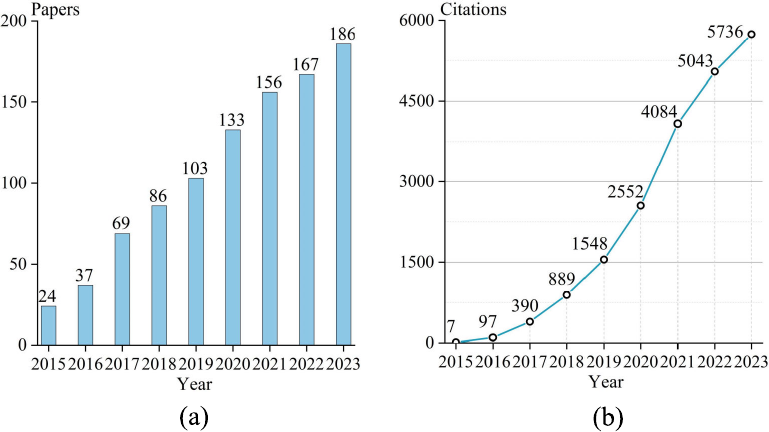}
\caption{The statistics of papers related to microstructure modeling (theories, algorithms, and applications): (a) number of annual papers; and (b) annual paper citations.}
\label{fig:trends}
\end{figure}

Geometric modeling is pivotal in the computer-aided design and manufacturing of products, and microstructures are no exception. It encompasses the representation (mathematical and/or computational models) and modeling algorithms (model creating, editing, and analyzing operations) that define and manipulate spatial information of microstructures. Because of microstructures' practical significance, research interests in this topic have seen a significant increase over the last decade, refer to the statistics~\footnote{The data were generated by searching the Web of Science website with combinations of keywords from the following three sets: ``\textit{geometric modeling} or \textit{representation} or \textit{creation}", ``\textit{cellular structures} or \textit{lattice structures} or \textit{porous structures} or \textit{truss structures}", ``\textit{additive manufacturing} or \textit{3d printing}".} in Fig.~\ref{fig:trends}.

Despite the increasing interest in microstructures, there remains a critical gap: the absence of a comprehensive review of geometric modeling methods for microstructure design and manufacturing. Most existing microstructure-related reviews reported research progress from the perspective of computational mechanics~\cite{2017_Dong_JMD,2022_review_cellular-materials,2023_reveiw_Thin-Walled-Structures}, design optimization~\cite{2020_Pan_ASB,2018_review_Feng_complex-topology-structures,2019_Nazir_IJAMT,2023_Noronha_review_Functionally-Graded-Lattices,2023_review_Lattice-Structured-Materials}, or manufacturing process~\cite{2018_Helou_lattice-design-manufacture,2022_review_lattice-with-SLM,2023_Borikar_review_progress-future-scope,2023_Almesmari_review_Polymeric-Metamaterials}. While some reviews touch on geometric aspects, such as those by Liu et al.~\cite{2021_Zhao_multiscale-lattice} and Letov et al.~\cite{2021_Letov_Challenges-and-Opportunities}, they focused on multiscale heterogeneous lattice structures and bio-inspired structures, two special kinds of microstructures.

\begin{figure}[t]
\centering
\includegraphics[width=0.5\textwidth]{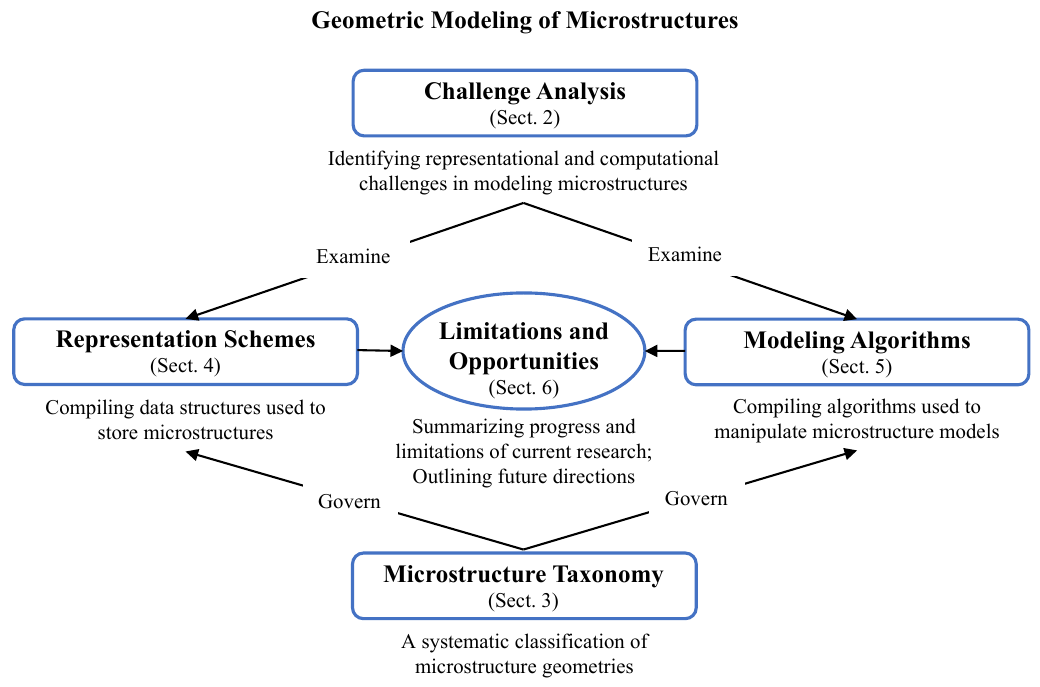}
\caption{Conceptual framework of this paper.}
\label{fig:paper-framework}
\end{figure}

This review provides a comprehensive, state-of-the-art literature review of geometric modeling methods used in microstructure design and manufacturing. As illustrated in Fig.~\ref{fig:paper-framework}, it aims to consolidate existing research studies into a coherent framework, covering topics of the challenges in microstructure modeling (Sect.~\ref{sec:challenges}), taxonomy of microstructures (Sect.~\ref{sec:taxonomy}), representation schemes (divided into topological and geometric categories, see Sect.~\ref{sec:representation}), modeling algorithms (categorized into design-oriented and manufacturing-oriented approaches, see Sect.~\ref{sec:operation}), as well as these methods' strengths and limitations (Sect.~\ref{sec:summary}). This review also discusses several promising research directions that may improve and/or complement current methods (Sect.~\ref{sec:summary}). It should be noted that this paper focuses on the fundamental geometric models and operations involved in microstructure design and manufacturing. It does not cover high-level applications such as design optimization or process planning, as there are already excellent reviews on those topics~\cite{2017_Dong_JMD,2022_review_cellular-materials,2023_reveiw_Thin-Walled-Structures,2020_Pan_ASB,2018_review_Feng_complex-topology-structures,2019_Nazir_IJAMT,2023_Noronha_review_Functionally-Graded-Lattices,2023_review_Lattice-Structured-Materials,2018_Helou_lattice-design-manufacture,2022_review_lattice-with-SLM,2023_Borikar_review_progress-future-scope,2023_Almesmari_review_Polymeric-Metamaterials}.

\section{Microstructure modeling challenges}
\label{sec:challenges}
AM holds the promise for several important disruptions: any geometry, any material, and any scale. However, as fabrication tools are advancing, computer-aided design (CAD) tools lag behind. This is partly due to the inability of conventional CAD model representations to keep pace with the geometric complexity achievable through AM, as will be discussed in Sect.~\ref{sec:gap-model-AM}. To bridge this gap, several critical challenges must be addressed, including representational compactness, computational efficiency, computational robustness, and multiscale integrity. These challenges will be detailed further in Sect.~\ref{sec:modeling-challenges}.

\subsection{The gaps between microstructure modeling and manufacturing}
\label{sec:gap-model-AM}
The capabilities of CAD tools have historically been shaped and limited by the way they represent shapes---how geometry is encoded and stored in computer memory~\cite{lipson2012design}. A particular representation will encourage or discourage certain types of objects and their manipulation. Most CAD systems today represent shapes by a dual scheme comprising boundary representation (B-rep) and constructive solid geometry (CSG)~\cite{zou2023variational}. B-rep describes geometry by its bounding surfaces, which are used to explicitly encode both intermediate and final geometry over the course of CAD modeling for internal references or downstream applications. CSG, on the other hand, describes geometry using combinations of primitive shapes, preserving the construction history of the geometry to aid in subsequent model editing (e.g., parametric editing).

While B-rep is precise and straightforward, it does not scale well to describe complex microstructures. This is because microstructures often have a high surface-area-to-volume ratio, which contrasts with B-rep's premise that storing boundaries is more compact than directly dealing with volumes~\cite{requicha1982solid}. As a result, a conventional CAD system would quickly run into limits of memory and computational power when modeling microstructures, leading to what is known as the model data explosion problem. For instance, it took the latest Siemens NX software (version 2306, Intel i5 Core, and 16GB memory) more than 4 hours to generate a commonly used lattice structure of 10mm$\times$10mm$\times$10mm volume and filled with repeating 0.1mm hex-star micro truss structures. Terrible lag also occurred when interacting with the model in the NX viewer window.

It should, however, be noted that whether or not a CAD system can handle microstructure modeling tasks depends on the number of microstructure elements a model has. Our experiments indicate that current CAD systems can handle simple microstructure models with hundreds or thousands of elements without issue. However, they struggle with complex microstructures containing millions of elements and are typically limited to handling only certain types, such as periodic microstructures.  Handling highly complex microstructures with billions of elements is beyond the capabilities of current CAD systems. For instance, a cube of 0.1m$^3$ volume with irregularly connected 0.1mm micro truss structures is estimated to require $6.4$TB of memory to instantiate all of its elements---roughly $10^{11}$ trusses and assuming 40 bytes (3 doubles for positions, and 2 doubles for length and radius) to store a single truss's geometric information and 24 bytes (6 ints for connection indices) to store its topological information. $6.4$TB memory is far beyond present CAD system capabilities for visualization, editing, analysis, etc. Such microstructures are increasingly common, exemplified by those cases in the DARPA TRADES Challenge Problems set~\cite{2023-TRADES-Challenge}.

\begin{figure}[t]
\centering
\includegraphics[width=0.5\textwidth]{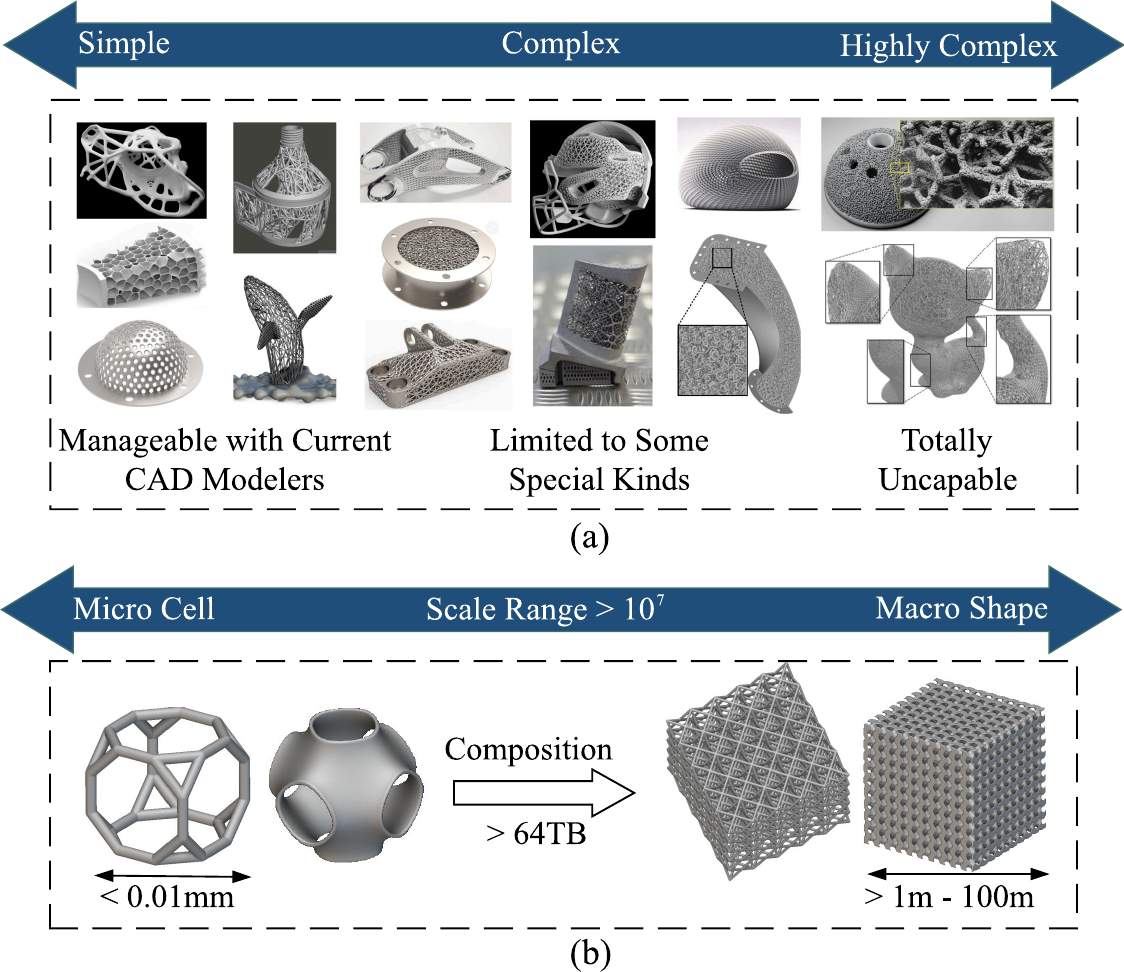}
\caption{Spectrum of microstructure complexity: (a) varying geometric complexity from simple to highly complex (adapted from George Allen's presentation~\cite{2016_siemens_george_allen}); and (b) multiscale geometric details.}
\label{fig:complexity}
\end{figure}

From the above discussions and examples, the gap between microstructure modeling and manufacturing primarily lies in managing models with millions or billions of elements, rather than simpler models with thousands. For notational simplicity, this paper categorizes microstructures at different complexity levels: simple structures (SS) for thousands of elements, complex structures (CS) for millions, and highly complex structures (HCS) for billions (see Fig.~\ref{fig:complexity} for examples).

Another important gap between microstructure modeling and manufacturing lies in microstructures' multiscale geometric details. At the micro-scale, intricate structures as small as approximately 0.01mm (continuing to decrease with advancing AM techniques) exist; at the meso-scale, there are structural cells and their connections; and at the macro-scale, there are large curvilinear surfaces and solids ranging from approximately 0.1m to 100m.  B-rep was not designed to manage such a hierarchical geometric complexity, let alone concurrently model shapes at different scales and ensure seamless interactions during model editing/optimization.

\subsection{The challenges of microstructure modeling}
\label{sec:modeling-challenges}
\begin{figure*}[t]
\centering
\includegraphics[width=0.84\textwidth]{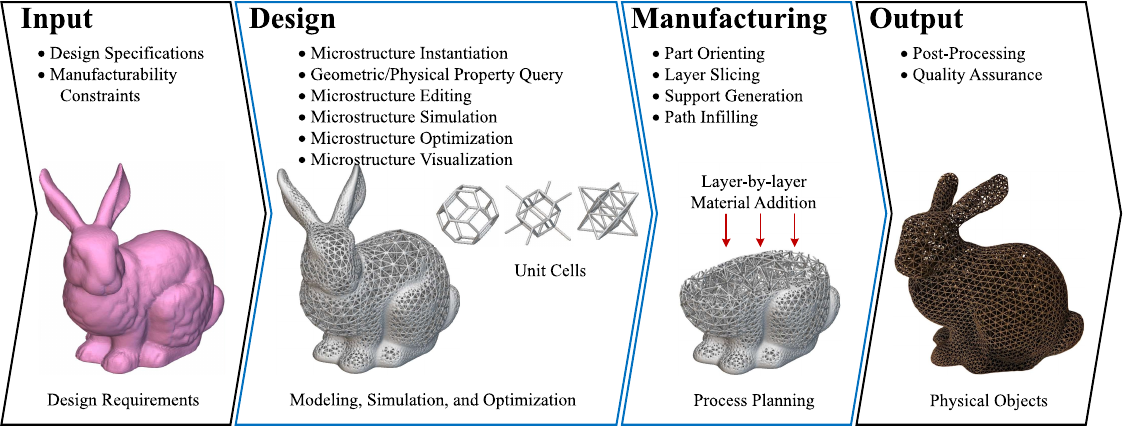}
\caption{Microstructure design and manufacturing pipeline.}
\label{fig:workflow}
\end{figure*}

The gaps identified earlier highlight the primary challenge in the geometric modeling of microstructures: efficiently storing and processing complex and highly complex microstructure (CS/HCS) models with limited computer memory, while maintaining acceptable performance. Additionally, when considering this problem within the broader context of microstructure design and manufacturing (Fig.~\ref{fig:workflow}), two more challenges emerge: robustness in processing the intricate geometric details of microstructures, particularly in tasks like slicing for AM tool paths, and ensuring consistency across geometric details at multiple scales when editing microstructure models.

\textbf{Representational compactness}. Except for some special kinds of microstructure models (e.g., periodically repeating or analytically describable), most CS/HCS models cannot be handled by current B-rep-based CAD systems due to the model data explosion problem. Developing a more compact representation scheme than B-rep is essential. 
On the one hand, according to CAD conventions, a CAD representation scheme must be informationally complete so that any possible geometric queries, e.g., point membership classification, from downstream applications can be automated~\cite{shapiro2002solid}.
On the other hand, a model being complete on geometric information tends to consume high computer memory, especially for CS/HCS models. To resolve this contradiction, we have to eliminate any redundant information (e.g., repeating topological and geometric patterns) in microstructure models. The challenge thus reduces to inventing highly compact geometric representation schemes for microstructures.

\textbf{Computational efficiency}. Editing a large CS/HCS model is time-consuming. 
First, because most model editing operations are local, there is an important culling step to attain an early rejection of microstructure elements that are irrelevant to a specific editing operation. The culling is usually done with the help of a spatial acceleration data structure, e.g., a K-D Tree. However, existing structures often fail to scale to CS/HCS models containing millions or billions of elements. Moreover, compute-intensive geometric computations, such as surface-surface intersections, pose additional challenges---there are still a large number of microstructure elements left after culling. The experiments presented in~\cite{2021_1D-skeleton_Zouqiang_convolution-surface} showed that an HCS model with 1 billion trusses involves an average number of $\sim500$ thousand trusses and a maximum number of $>1$ million trusses in a single slicing operation. 
Therefore, efficient acceleration data structures and fast geometric algorithms capable of handling large-scale microstructures are crucial.

\textbf{Computational robustness}. The complexity of microstructure geometries leads to frequent degenerate cases like tangencies and overlaps during editing operations, which are likely to occur more often than what we have in today's CAD systems. Also, many degenerated cases occurring at the same time make it easy to form coupled degeneracies, adding more complexity to the detection and resolution of degenerated cases. These two lead to the third challenge of microstructure modeling: robustness toward processing intricate geometric details of microstructures.

\begin{figure}[t]
\centering
\includegraphics[width=0.49\textwidth]{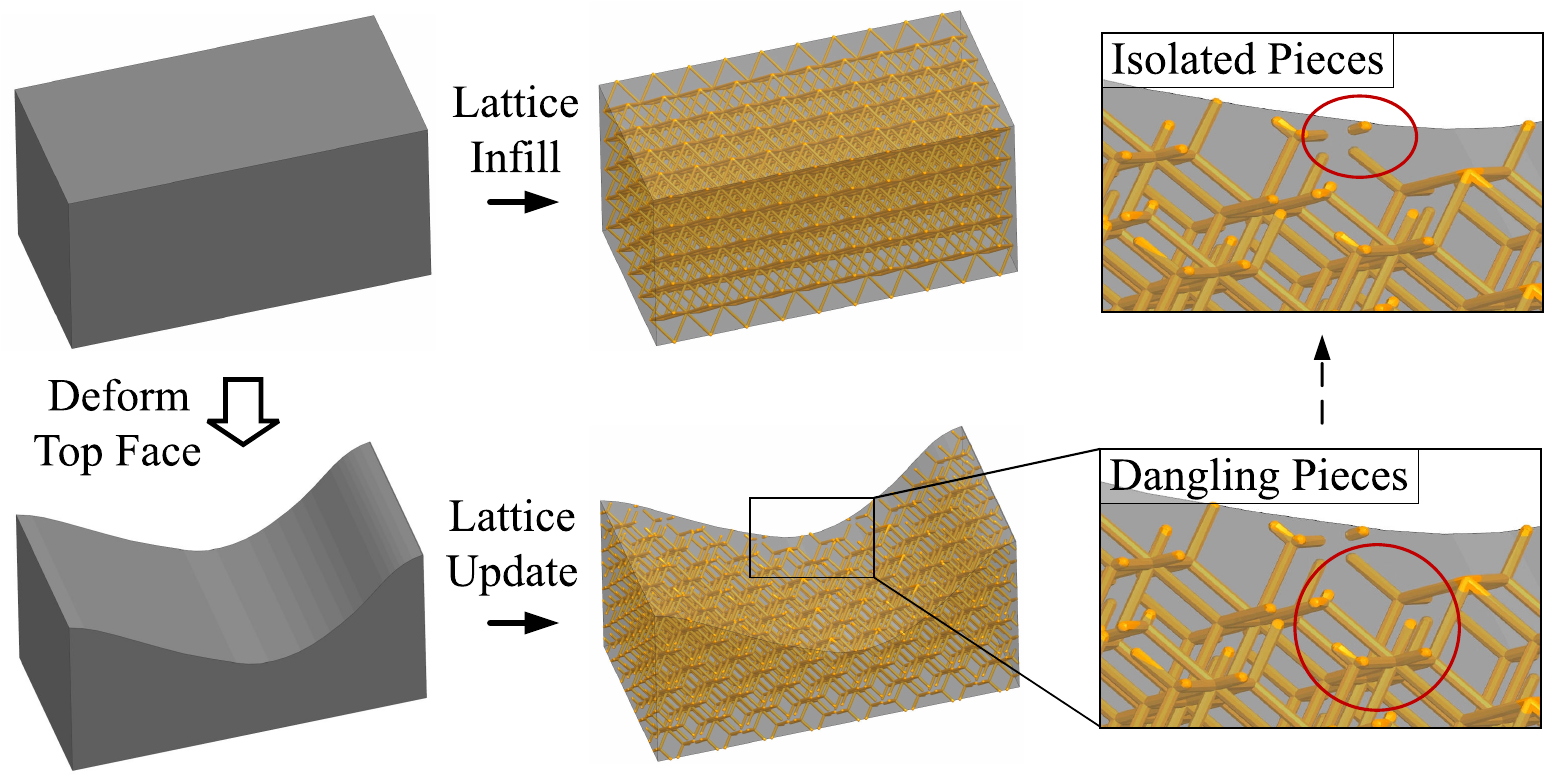}
\caption{The modeling consistency issue of Siemens NX.}
\label{fig:consistency}
\end{figure}

\textbf{Multiscale integrity}. As already noted, microstructures exhibit geometric details across multiple scales, from micro to macro. Maintaining consistency in these geometric details is crucial. A typical example is shape conformity; if micro-level structures are generated independently of the macro-level boundary shape, it can lead to reduced performance or even create physically impossible microstructure models.
Furthermore, when geometric details at one scale are modified, those changes will not automatically propagate to other scales. As a result, the integrity across multiple scales is broken, and an invalid or unintended microstructure model is generated. 
For example, assuming there is a boundary-conformed microstructure model (see the examples at the top row of Fig.~\ref{fig:taxonomy}) and we want to deform its macro boundary shape; without an automatic change propagation mechanism among geometric details at different scales, micro-scale structures will remain unchanged, and their trimming with the new macro boundary could produce dangling pieces and even isolated pieces (see Fig.~\ref{fig:consistency}), resulting in invalid, physically impossible models. 
Current CAD systems often lack mechanisms for automatic change propagation among multiscale geometric details, posing challenges in maintaining integrity during microstructure editing.

It is important to note that the challenges discussed here represent some of the primary issues in microstructure modeling. Other challenges certainly exist and should be considered for specific modeling contexts and practices. For example, controlling microstructure parameters is both important and challenging for practical applications. A typical application is the gradual infill generation for 3D printing, where higher density is concentrated near the boundary and lower density is used in the interior. This involves advanced density design and topology optimization methods, both of which pose significant challenges.

\begin{figure*}[b]
\centering
\includegraphics[width=0.8\textwidth]{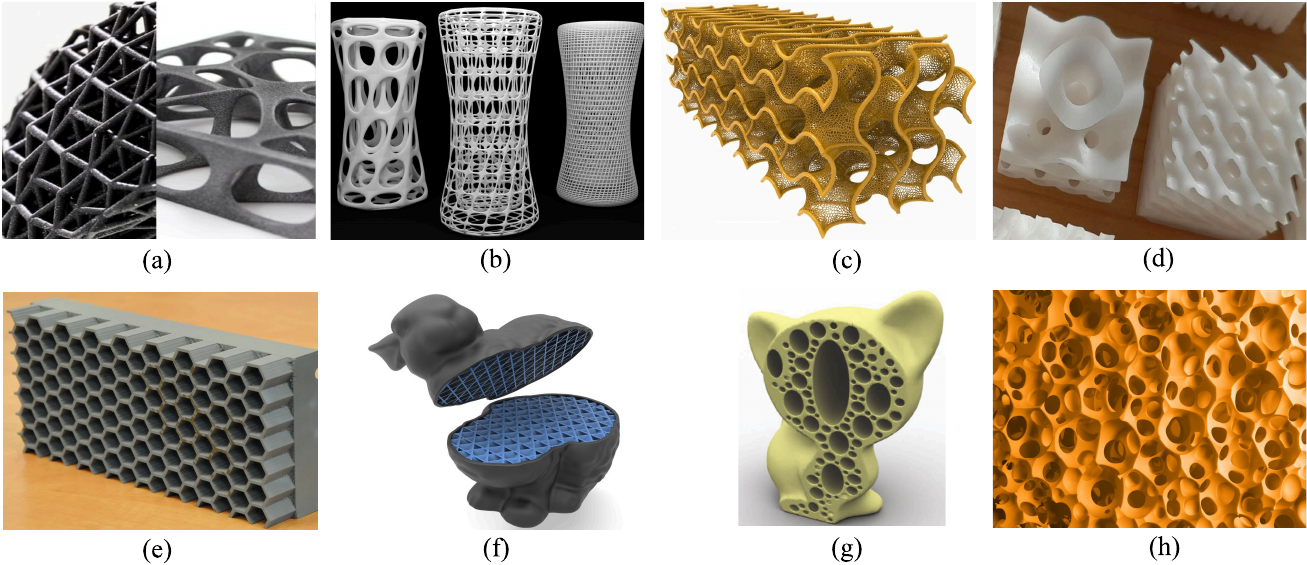}
\caption{Microstructure examples: (a)(b) lattice structures; (c)(d) TPMS structures; (e)(f) shell-based structures; and (g)(h) foam structures ((g) is from~\cite{2018_support_Lee_self-support-infill_ellipse-hollowing}).}
\label{fig:typical_microstructure}
\end{figure*}

\section{Microstructure taxonomy}
\label{sec:taxonomy}
The term ``microstructure" encompasses a diverse array of structural forms, e.g., lattice structures, triply periodic minimal surface (TPMS) structures, and foam structures (see Fig.~\ref{fig:typical_microstructure} for examples), among others. Systematically classifying these microstructures is crucial not only for deepening our understanding but also for comprehensively reviewing current research progress. While existing classifications have primarily focused on specific types such as lattice structures~\cite{2017_Dong_JMD,2020_Pan_ASB,2018_Savio_ABB,2018_Tamburrino_JCISE,2019_Nazir_IJAMT}, this section presents a unified three-level taxonomy encompassing all microstructure types, depicted in Fig.~\ref{fig:taxonomy}.

\subsection{Macro-level classification by boundary shapes}
\label{sec:macro-level}
\begin{figure*}[t]
\centering
\includegraphics[width=0.8\textwidth]{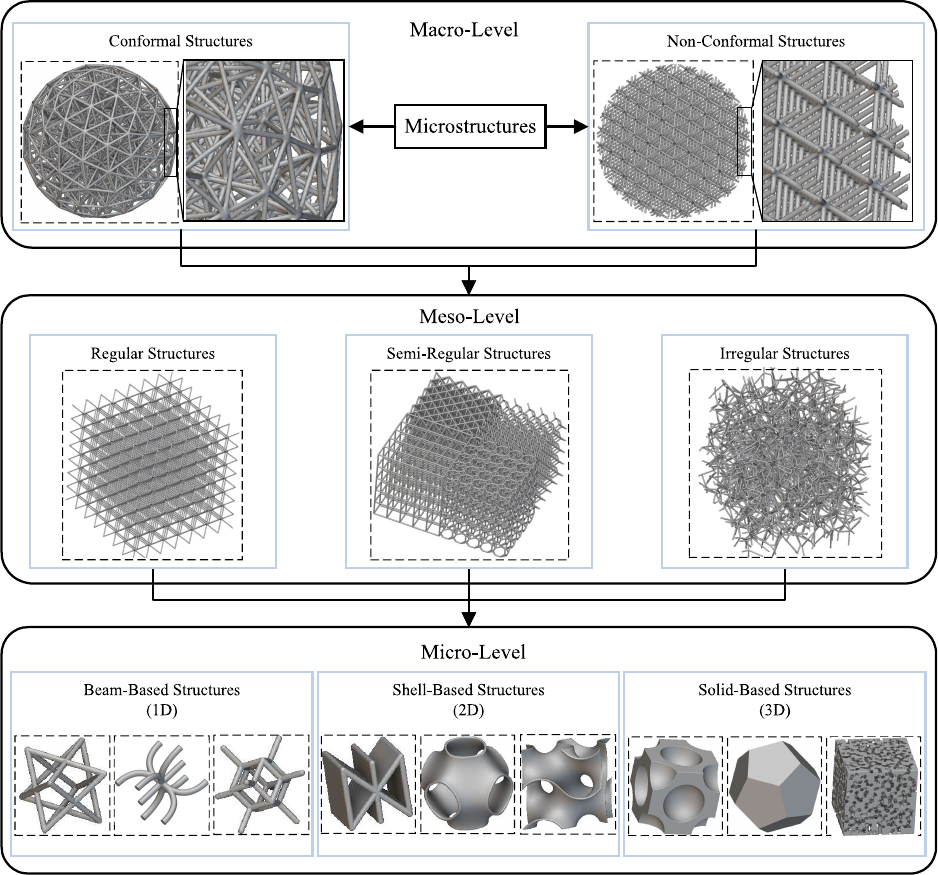}
\caption{The three-level taxonomy of microstructures.}
\label{fig:taxonomy}
\end{figure*}

At this level, microstructures are simply classified based on their adherence to the overall boundary shape. Microstructures that closely follow the boundary shape are classified as boundary-conformed, illustrated in the top row of Fig.~\ref{fig:taxonomy}. Conversely, microstructures that do not conform to the boundary shape are classified as non-conformal.

Boundary-conformed microstructures typically require careful design and optimization processes such as parametrization~\cite{2018_3D-rep_Massarwi_trivariate-spline_extension-Hierarchical}, hex-meshing~\cite{2012_conformal_volumetric-meshing,2024_conformal_TPMS+lattice}, or sphere packing~\cite{2021_spherePacking_conformal}. These optimizations need to consider the entire microstructure globally rather than focusing solely on micro-level structures near the boundary. This is because changes made to one micro-level structure affect its neighboring structures due to associativity, i.e., propagating local changes outward. 
In this regard, while boundary-conformed microstructures offer good mechanical properties, achieving this conformity often comes at a high price.

On the other hand, non-conformal microstructures are generated by directly trimming micro-level structures with the macro-level boundary, as depicted in the top row of Fig.~\ref{fig:taxonomy}. This approach is computationally simple but can result in microstructures with dangling and/or isolated pieces. Such configurations typically exhibit reduced mechanical properties, e.g., strength, and can sometimes yield physically impossible microstructures.

\subsection{Meso-level classification by topological regularity}
\label{sec:meso-level}
Regardless of whether a microstructure is boundary-conformed, its micro-level structures can be arranged in various ways. The simplest arrangement involves repeating a cell of micro-level structures along the x-, y-, and z-dimensions, resulting in a regular microstructure type (see the mid-row of Fig.~\ref{fig:taxonomy}, left). At the opposite end of the spectrum, the most complex arrangement exhibits no topological regularity among the micro-level structures (see the mid-row of Fig.~\ref{fig:taxonomy}, right), defining the irregular type~\cite{2022_Groth_stochastic-lattice-creation,2016_polyhedral-voronoi-scaffolds}. Between these extremes lie semi-regular microstructures, where the topology exhibits local periodicity or follows predefined variation patterns, such as functionally graded lattice structures~\cite{2021_1D-skeleton_Zouqiang_convolution-surface}.

Regular microstructures are straightforward to model but offer limited mechanical properties. Irregular microstructures, on the other hand, provide a wide range of mechanical properties and are highly versatile, but their representational and computational efficiency may pose practical challenges (i.e., the need to capture informational completeness and stochastic topology of irregular microstructures leave a narrow space for data compression and computing acceleration). Semi-regular microstructures strike a balance, offering diverse mechanical properties while benefiting from local periodicity or predefined variation patterns for more compact representations and efficient computations.

In summary, while regular microstructures are easy to handle and irregular microstructures offer broad applicability, semi-regular microstructures present a promising middle ground by balancing mechanical diversity with computational feasibility. It should, however, be noted that editing semi-regular microstructures often leads to transitions into irregular forms due to the inherent locality and variability of model edits.

\subsection{Micro-level classification by cell geometries}
\label{sec:micro-level}
At the very bottom level of the taxonomy, the focus shifts to the specific geometry of micro-level structures, called cell geometries in this work. As depicted in Fig.~\ref{fig:taxonomy}, cell geometries exhibit considerable diversity and can undergo transformations by varying their control sizes, thereby not trivially distinguishable. To classify these geometries effectively, we adopt a dimension-based approach that disregards size influence. Cell geometries are categorized into three types: beam-based, shell-based, and solid-based. This classification draws inspiration from solid mechanics, which categorizes objects into 1D elements (rods, beams), 2D elements (plates, shells), and 3D elements (solids). Likely, beam-based cell geometries consist of 1D elements, either straight or curved, connected at their ends. Shell-based cell geometries represent thickened surfaces, while solid-based cell geometries encompass solid volumes with small internal features (blind or through holes, regular or irregular).

The above dimension-based classification does not carry any size information. Adding sizes introduces another variant: uniform (homogeneous) microstructures and non-uniform (heterogeneous) microstructures. In uniform microstructures, all cell geometries share identical size parameters, typically resulting in isotropic mechanical properties. In contrast, non-uniform microstructures exhibit spatially varying size parameters within or across cells, leading to anisotropic mechanical properties. For instance, density-graded microstructures can be generated by varying beam radii in a topologically regular beam-based lattice structure.

This hierarchical classification approach---from macro-level boundary shapes, to meso-level topologies, and finally to micro-level cell geometries---provides a systematic framework for understanding and categorizing microstructures. This taxonomy serves as a foundation for compiling literature on microstructure modeling, as well as guiding the review of microstructure representation schemes in Sect.~\ref{sec:representation}, and modeling algorithms in Sect.~\ref{sec:operation}.

\section{Microstructure representation schemes}
\label{sec:representation}
Building upon the hierarchical taxonomy of microstructures established earlier, effective representation schemes must address three key hierarchical aspects: macro-level boundary shape, meso-level topology, and micro-level cell geometry. While conventional CAD representation methods suffice for macro-level boundary shapes, specialized approaches are needed for capturing the intricacies of meso-level topologies and micro-level cell geometries. This section focuses on existing topological representation methods (Sect.~\ref{sec:topological-rep}) and geometric representation methods (Sect.~\ref{sec:geometric-rep}), highlighting their roles in accurately and efficiently capturing microstructure details.

\subsection{Topological representations}
\label{sec:topological-rep}
Topological representations refer to the mathematical models and data structures used to describe and store the arrangement information of microstructures. In its most general form, the arrangement can be modeled with graphs. Nevertheless, the regularity in some microstructures allows for more compact representations. The following discussions present these variations in representation compactness in detail.

\subsubsection{Regular topology}
\label{sec:regular}
Regular topology refers to a periodic, repeating arrangement of cell geometries, as illustrated in the mid-row of Fig.~\ref{fig:taxonomy}. To explicitly represent this arrangement pattern, a 3D grid, which is a specialized class of graph, can be employed. In this representation, nodes encode individual cells, while edges describe their directional repetitions. However, such a representation inherently contains redundancies since cells do not vary by location within the structure.

\begin{figure}[t]
\centering
\includegraphics[width=0.48\textwidth]{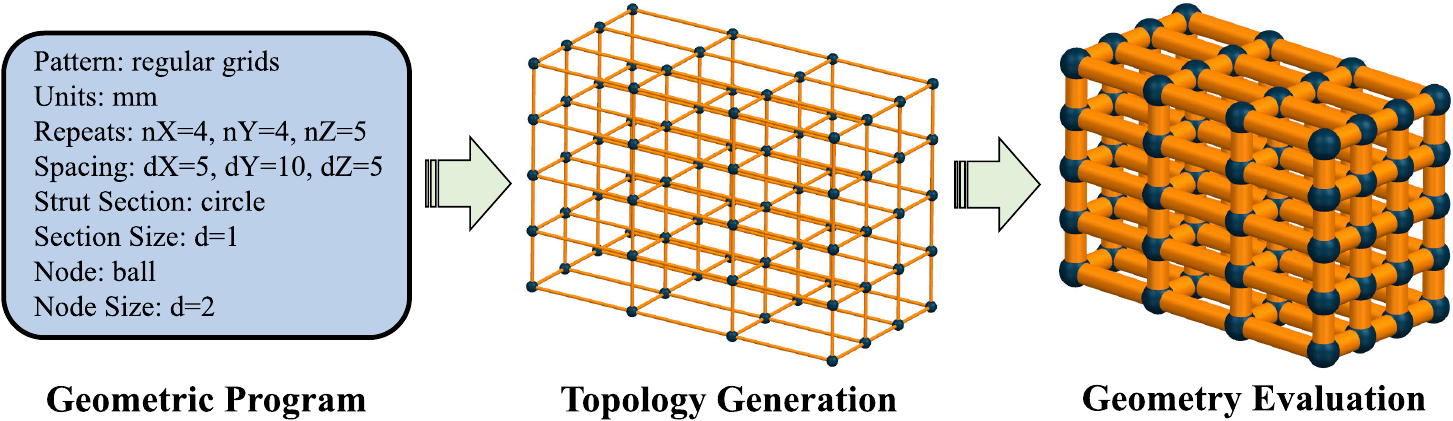}
\caption{Example of program-based regular microstructure representation.}
\label{fig:regular_pattern}
\end{figure}

In 2016, George Allen~\cite{2016_siemens_george_allen} suggested the idea of using a procedural approach to address these redundancies, which was subsequently implemented by Gupta et al.~\cite{2019_3D-rep_Gupta_CSG_steady-lattice}. This procedural method describes microstructures as a construction process or a geometric program, akin to structured programming languages (see Fig.~\ref{fig:regular_pattern}). Using this approach, storing a lattice composed of one million repeating cells would require only a few program statements, consuming less than 1KB of memory.

\textbf{Remarks:} While the program-based method effectively handles regular microstructures with efficiency, it falls short in scenarios requiring model edits. Changes to program statements propagate globally across the entire microstructure, limiting the ability to make localized adjustments crucial for varying shapes and mechanical properties in microstructure modeling.

\subsubsection{Semi-regular topology}
\label{sec:semi-regular}
\begin{figure}[t]
\centering
\includegraphics[width=0.4\textwidth]{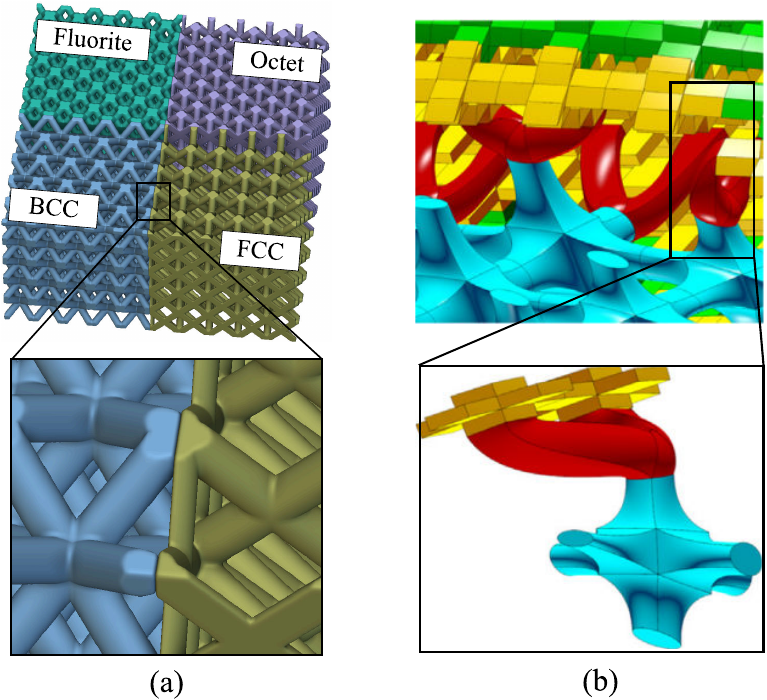}
\caption{Semi-regular topology transition: (a) natural transition; and (b) topology adaption (the red trusses)~\cite{2021_3D-rep_Hong_trivariate-spline_Trimmed-Trivariate}.}
\label{fig:topo_combination}
\end{figure}

Semi-regular topology refers to microstructures where the arrangement of cells exhibits local periodicity or follows a predefined variation pattern. In microstructures with local periodicity, multiple regular sub-microstructures coexist, each with its own topology but capable of interfacing naturally with others, as depicted in Fig.~\ref{fig:topo_combination}a. To represent such microstructures, existing methods typically extend regular microstructure representation schemes by incorporating occupancy information for each sub-microstructure~\cite{2020_topology-rep_Liu_Multi-Topology-Lattice,2019_1D-skeleton_Leonardi_multi-topology}. Moreover, the representation is associated with a library of primitive cells (in the format of construction process/programs) and a placement record of these cells (i.e., the occupancy information).

In cases where sub-microstructures cannot naturally interface, specialized topology transition methods become necessary. Three strategies have been proposed. The first strategy utilizes implicit representations (such as distance functions, sigmoid functions, and beta growth functions) to blend different topologies~\cite{2015_topology-rep_Yoo_beta-growth-function,2023_1D-skeleton_Letov_topology-transition,2021_MSLattice_multi-topology-implicit,2020_procedural_freely-orientable}. The second strategy leverages modern data-driven techniques~\cite{2020_data-driven_shape-DNA,2020_data-driven_deep-generative,2020_spinodoid-metamaterials,2023_Ha_data-driven-AI,2023_Zheng_data-driven-generative,2015_data-driven_Control-Elasticity,2017_data-driven-two-scale,2024_data-driven_review}, which begin by generating a dataset of unit cells with varying properties, then select and assemble the optimal ones from the dataset while ensuring cell boundary compatibility. For such methods, the structure-property relationship is essential and often modeled using neural networks or optimization algorithms. The quality of datasets is equally important, and Lee et al.~\cite{2023_Lee_t-metaset} has presented a typical method to generate high-quality datasets based on active learning. Snapp et al.~\cite{2024_SDL_energy-absorb} extended this work and utilized active learning to optimize energy-absorbing structures. Microstructures designed using the above data-driven methods exhibit a larger property space than those based on periodic arrangements. This is especially true for non-periodic spinodoid metamaterials~\cite{2020_spinodoid-metamaterials}, where their simple yet effective parametrization enables a rich and continuously tunable property space. The last strategy adds a topology adaptation layer to directly connect disparate topologies (see Fig.~\ref{fig:topo_combination}b)~\cite{2021_3D-rep_Hong_trivariate-spline_Trimmed-Trivariate}. While conceptually straightforward, these adaptation algorithms are complex, often resulting in highly distorted connecting elements (e.g., the red trusses in Fig.~\ref{fig:topo_combination}b).

The other type of semi-regular topology involves topological variations following predefined patterns. These patterns are currently defined using scalar, vector, or tensor fields. Scalar fields include distance fields~\cite{biswas2004heterogeneous} and density fields~\cite{2021_1D-skeleton_Zouqiang_convolution-surface}; vector fields are represented by volumetric B-splines~\cite{2017_3D-rep_Elber_trivariate-spline_Functional-Composition}; and tensor fields include elasticity tensor fields generated, for example, through topology optimization methods~\cite{2022_Maurizi_inverse-design-lattice}.

\textbf{Remarks:} Various effective methods exist for representing semi-regular topology, addressing challenges such as topology transition. Although presented in different forms, the idea common to most existing methods is storing how the topology varies, not the final topology directly. However, they currently lack a unified approach and often provide limited flexibility in model editing capabilities.

\subsubsection{Irregular topology}
\label{sec:irregular}
Unlike regular and semi-regular counterparts, irregular microstructures exhibit stochastic topologies without explicit patterns. Consequently, most existing research studies store the final topology directly as a graph, using adjacency matrices, adjacency lists, or edge lists~\cite{2017_light-weight-triangulation,2019_hybrid_tang}.

Graph-based methods may suffice for the moment because irregular microstructures are often generated by topology optimization that typically yields small-scale irregular microstructures~\cite{2023_support_Wang_self-support-deformation_subdivision-simplicification}. However, as techniques for irregular microstructure generation advance, scaling up to millions or billions of elements is becoming feasible~\cite{2022_Groth_stochastic-lattice-creation,zou2024meta}. For instance, Voronoi-based methods are capable of generating large-scale irregular microstructures~\cite{2016_polyhedral-voronoi-scaffolds,2016_topology-rep_Fantini_Voronoi-lattices,2020_topology-rep_Lei_Parametric-design-Voronoi-based,2016_procedural-voronoi}, surpassing the capabilities of graph-based representations. Particularly, Martinez et al.~\cite{2016_procedural-voronoi} presented a procedural Voronoi-based approach to generate large-scale irregular microstructures with varying mechanical properties. To address the memory bottlenecks when processing large-scale microstructures, recent studies by Liu et al.~\cite{2021_1D-skeleton_Zouqiang_convolution-surface} and Zou et al.~\cite{zou2024meta} have introduced streaming-based computing techniques and decomposition-recombination plans into the modeling of large-scale irregular microstructures.

\textbf{Remarks:} The primary challenge in representing irregular topologies lies in scaling up to handle large-scale irregular microstructures (CS/HCS), rather than small-scale ones. Current research is still in its early stages, and the major research problems include compact representations, efficient topological queries, and real-time editing capabilities tailored to these complex structures.

\def\tabularxcolumn#1{m{#1}}
\begin{table*}[htbp]
\caption{Microstructure representation schemes: Lines of development.}
\centering
\setlength\extrarowheight{8pt}
\begin{tabularx}{\textwidth}{
    |>{\centering\arraybackslash}>{\hsize=.3\hsize\linewidth=\hsize}X
    |>{\centering\arraybackslash}>{\hsize=1.1\hsize\linewidth=\hsize}X
    |>{\centering\arraybackslash}>{\hsize=1.1\hsize\linewidth=\hsize}X
    |>{\centering\arraybackslash}>{\hsize=1.1 \hsize\linewidth=\hsize}X 
    |>{\centering\arraybackslash}>{\hsize=1.4\hsize\linewidth=\hsize}X 
    |>{\centering\arraybackslash}>{\hsize=\hsize\linewidth=\hsize}X|
   }
    \hline
    \multirow{2}{*}{Time}& \multicolumn{5}{>{\hsize=\dimexpr5\hsize+5\tabcolsep+\arrayrulewidth\relax}>{\centering\arraybackslash}X|}{Geometric Representations} \\
    \cline{2-6}
    &1D Curved-Based &2D Surface-Based &3D Volume-Based &3D Implicits-Based &Hybrid\\
    \hline
    2000 &  &Generating lattice by stacking STL trusses~\cite{2005_HongqingWang_2Dsurface,2002_HongqingWang_conformal} & & & \\

    2005 & Lattices with curved axes~\cite{2005_1D_curve-axis_Huang_curved-cell-edges}; approximated nodal geometry~\cite{2005_1D_convex-hull_solidifying-wireframe}
    &Surface mesh-based lattices~\cite{2006_chenYong_mesh-based} 
    & 3D texturing-based lattices~\cite{2006_chenYong_mesh-based,2007_ChenYong_3DtextureMapping}
    & 
    & Combination of B-rep and STL~\cite{2005_HongqingWang_2Dsurface}\\
    
    2010 & Parametric lattices~\cite{2008_1D-skeleton_Chenchu_DFAM} 
    & 3D texture mapping-based lattices~\cite{2007_ChenYong_3DtextureMapping} 
    & FVM-based microstructures~\cite{2013_3D-rep_Feng_FVM_macromolecules}; mapping-based TPMS structrues~\cite{2012_Yoo_voxelMapping,2011_Yoo_hexMapping}
    & F-rep-based lattices~\cite{2011_Pasko_FrepLattice,2013_4D-rep_Fryazinov_Frep_multi-scale} 
    &\\

    2015 & Blended nodal geometry~\cite{2017_1D_Panetta_blended-nodes_Stress-Relief} 
    &Stacking prefabricated STL trusses~\cite{2017_prefab-cell_HGM}; faceted diamond-based STL lattices~\cite{2017_factedDiamondMapping}; lightweight triangulation of lattices~\cite{2017_light-weight-triangulation}
    &Voxel-based lattices~\cite{2017_3D-rep_Aremu_voxel_trimmed-lattice}; spline-based microstructures~\cite{2017_3D-rep_Elber_trivariate-spline_Functional-Composition,2018_3D-rep_Massarwi_trivariate-spline_extension-Hierarchical} 
    & 
    &Combination of trimmed B-rep and STL~\cite{2017_prefab-cell_HGM}\\

    2020 & VDF-based lattices~\cite{2020_1D-skeleton_Liang_VDF-conformal}; blended nodes~\cite{2021_1D-skeleton_Azman_different-types-of-joint,2019_1D-skeleton_Tang_convolution-surface,2021_1D-skeleton_Zouqiang_convolution-surface}; spline-based axes~\cite{2021_3D-rep_Hong_trivariate-spline_Trimmed-Trivariate}; lattice with curved axes~\cite{2018_1D_non-planar-curved-lattice,2021_1D_curve-axis_Bai_curved-lattice-strut,2021_1D_curve-axis_Fu_curved-beam,2021_1D_curve-axis_Feng_curved-cell-walls,2022_1D_curve-axis_Cao_curved-microstructure}; lattice with quadric/freeform profiles~\cite{2018_quador,2019_hybrid_Gupta_CSG+CST+Brep,2021_slicing_efficiency_matrix-oriented-data-structure,2018_1D_curve-profile_Zhang_free-form-truss}; approximated nodal geometry with convex hull~\cite{2020_CHoCC,2020_convexHull}
    &Parametric surface-based lattices~\cite{2021_spherePacking_conformal}; subdivision surface-based lattices~\cite{2018_subdivision_2Dsurface,2019_differentialOffsetGrading} 
    &CSG-based lattices~\cite{2019_3D-rep_Gupta_CSG_steady-lattice} ; bio-inspired volumetric microstructures~\cite{2020_3D-rep_letov_bio-inspired_volumetric-cell}; spline-based volumetric microstructures~\cite{2021_3D-rep_Hong_trivariate-spline_Trimmed-Trivariate}; voronoi-based microstructures~\cite{2020_topology-rep_Lei_Parametric-design-Voronoi-based,2022_polyhedral-voronoi-cancellous}; parameterized hexmesh-based microstructures~\cite{2018_2Dsurface_chen_mapping-deformation}
    & TPMS-based porous structures~\cite{2022_4D-rep_Wang_TPMS}; F-rep and STL-free microstructures~\cite{ding2021stl} 
    & Combination of CSG, Brep, and CST~\cite{2019_hybrid_Gupta_CSG+CST+Brep}\\

     2025 & Lattices with deformed axis~\cite{2024_1D-rep_curved-axis_deformation-control}; curved G-lattice~\cite{2023_1D_curve-axis_Armanfar_G-lattice}
     & Meta-mesh-based lattices~\cite{zou2024meta}; subdivision surface-based lattices~\cite{2023_2Dsurface_Xiong_Subdivisional-modelling}
     & 
     & F-rep and skeleton-based microstructures~\cite{2024_F-rep_zhao}
     & \\
    \hline
\end{tabularx}%
\label{tab:line-of-representation}%
\end{table*}%

\subsection{Geometrical representations}
\label{sec:geometric-rep}
To be aligned with the dimension-based taxonomy outlined in Sect.~\ref{sec:micro-level}, existing methods for geometric representation are categorized based on their dimensional aspects. Table~\ref{tab:line-of-representation} provides an overview of some representative methods in use today, and the following discussions provide details.

\subsubsection{1D curve-based representations}
\label{sec:1D}
This type of geometric representation is tailored for beam-based microstructures. The idea is simple: augmenting the topological representation schemes presented in Sect.~\ref{sec:topological-rep} with parameters describing the geometry of beams, such as axis shape, radius, and length~\cite{2019_3D-rep_Gupta_CSG_steady-lattice,2017_light-weight-triangulation,2008_1D-skeleton_Chenchu_DFAM,2019_hybrid_tang,2020_1D-skeleton_Liang_VDF-conformal,2019_1D-skeleton_Tang_convolution-surface,2021_1D-skeleton_Zouqiang_convolution-surface, 2022_1D-skeleton_Letov_Frep_beamTopology,2020_CHoCC,2020_convexHull,2021_1D-skeleton_Azman_different-types-of-joint,2022_1D_curve-axis_Cao_curved-microstructure,2021_1D_curve-axis_Bai_curved-lattice-strut,2005_1D_curve-axis_Huang_curved-cell-edges,2021_1D_curve-axis_Fu_curved-beam,2021_1D_curve-axis_Feng_curved-cell-walls,2021_3D-rep_Hong_trivariate-spline_Trimmed-Trivariate,2023_1D_curve-axis_Armanfar_G-lattice,2018_quador,2019_hybrid_Gupta_CSG+CST+Brep,2018_1D_curve-profile_Zhang_free-form-truss,2021_slicing_efficiency_matrix-oriented-data-structure,zou2024meta}. Depending on the topological regularity of a microstructure, the augmentation can be done by adding new program statements about the geometric parameters (for regular topology)~\cite{2019_3D-rep_Gupta_CSG_steady-lattice}, or by transforming the geometric parameters into vector fields (for semi-regular topology)~\cite{2022_1D-skeleton_Letov_Frep_beamTopology}, or by adding the geometric parameters to graphs as their weights (for irregular topology)~\cite{zou2024meta}.

\begin{figure}[h]
    \centering
    \includegraphics[width=0.475\textwidth]{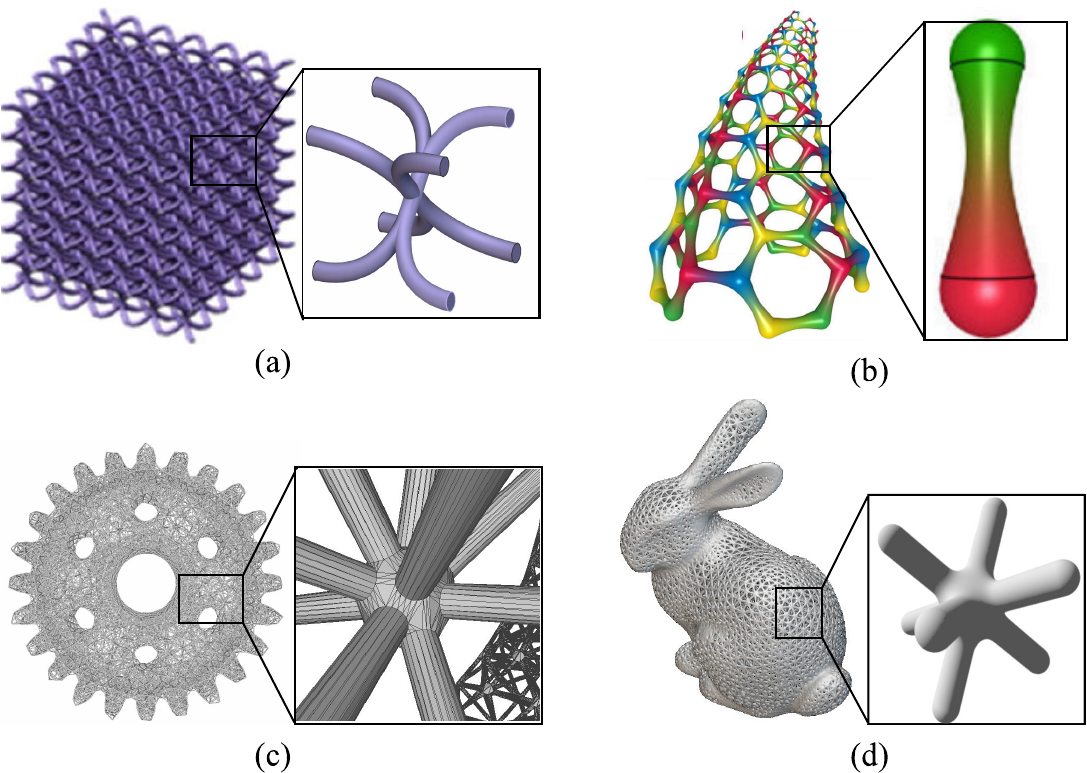}
    \caption{Beam shape variants: (a) axis-curved beams~\cite{2021_1D_curve-axis_Bai_curved-lattice-strut}; (b) profile-curved beams~\cite{2018_quador}; (c) convex hull nodes~\cite{2020_convexHull}; and (d) blended nodes~\cite{2021_1D-skeleton_Zouqiang_convolution-surface}.}
    \label{fig:joints}
\end{figure}

There have been significant advancements in enhancing these augmented representation schemes. One line of research focuses on extending the beam shape to increase design freedom in two directions:
\begin{enumerate}
    \item Transitioning from straight lines to curved axes, such as arcs~\cite{2024_1D-rep_curved-axis_deformation-control,2022_1D_curve-axis_Cao_curved-microstructure,2021_1D_curve-axis_Bai_curved-lattice-strut,2021_1D_curve-axis_Fu_curved-beam,2021_1D_curve-axis_Feng_curved-cell-walls,2021_3D-rep_Hong_trivariate-spline_Trimmed-Trivariate,2023_1D_curve-axis_Armanfar_G-lattice,2005_1D_curve-axis_Huang_curved-cell-edges,2018_1D_non-planar-curved-lattice}, as shown in Fig.~\ref{fig:joints}a; and
    \item Extending from linear profiles to curved profiles, such as quadratic curves~\cite{2018_quador,2019_hybrid_Gupta_CSG+CST+Brep,2018_1D_curve-profile_Zhang_free-form-truss,2021_slicing_efficiency_matrix-oriented-data-structure}, as shown in Fig.~\ref{fig:joints}b.
\end{enumerate}

Shifting from a line to a curve is a simple concept to express, but its implementation presents significant challenges, especially regarding the associated computational overhead. For complex microstructures (CS/HCS), any additional computation at the micro-level scales exponentially across millions or billions of elements.

The other line of research focuses on enhancing the shape of junction nodes, again, in two directions:
\begin{enumerate}
    \item Replacing spherical nodes with their approximations for improved efficiency and robustness in geometric calculations~\cite{2020_CHoCC,2020_convexHull,2005_1D_convex-hull_solidifying-wireframe}, as depicted in Fig.~\ref{fig:joints}c; and
    \item Substituting spherical nodes with their blended variants to enhance mechanical properties and reduce stress concentration~\cite{2019_1D-skeleton_Tang_convolution-surface,2021_1D-skeleton_Zouqiang_convolution-surface,2017_1D_Panetta_blended-nodes_Stress-Relief}, as depicted in Fig.~\ref{fig:joints}d.
\end{enumerate}

Recent work by Makatura et al.~\cite{2023_Procedural-Metamaterials_curve-beam} even demonstrates that augmented representation schemes can extend beyond beam-based microstructures to include shell-based and solid-based types by enhancing graphs with construction procedures and spatially varying thicknesses.

\textbf{Remarks:} 1D curve-based representation schemes have been extensively researched, resulting in numerous available algorithms. These algorithms are effective for most small-scale microstructure modeling scenarios. However, further development is needed for complex microstructures (CS/HCS), particularly in addressing the challenge of managing computational overheads associated with large-scale models.

\subsubsection{2D surface-based representations}
\label{sec:2D}
This category of methods relies on B-rep to directly store the boundaries (2D manifolds/surfaces) of cell geometries. Unlike 1D curve-based methods tailored for beam-based microstructures, 2D surface-based methods are versatile and applicable to all types of microstructures due to the generic nature of B-rep. Conventional B-rep schemes, such as mesh surfaces~\cite{2005_HongqingWang_2Dsurface,2002_HongqingWang_conformal,2017_prefab-cell_HGM,2006_chenYong_mesh-based,2015_blending-mesh-parametric-surface}, parametric surfaces~\cite{2021_spherePacking_conformal}, and subdivision surfaces~\cite{2023_2Dsurface_Xiong_Subdivisional-modelling,2018_subdivision_2Dsurface,2019_differentialOffsetGrading}, have all been used for this purpose, as detailed below.

\begin{figure*}[t]
\centering
\includegraphics[width=0.7\textwidth]{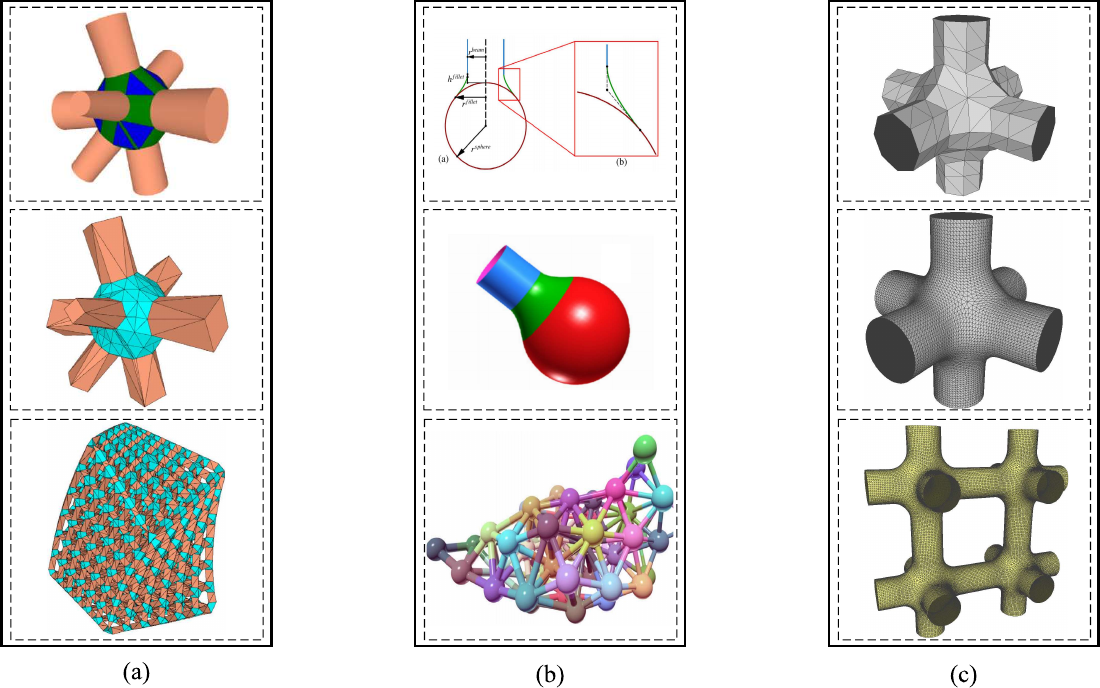}
\caption{Illustrations of surface-based representations: (a) mesh-based lattice structures~\cite{2020_CHoCC}; (b) parametric surface-based lattice structure~\cite{2021_spherePacking_conformal,zou2024tpms2step}; and (c) subdivision surface-based lattice structure~\cite{2018_subdivision_2Dsurface,2019_differentialOffsetGrading}.}
\label{fig:surface}
\end{figure*}

\textbf{Mesh-based methods.} 
Most studies in this category focus on mesh generation for cell geometries, primarily for beam-based microstructures. Their common focus is generating high-quality mesh models of cell geometries. The quality criteria can be geometric (e.g., discretization errors~\cite{2020_CHoCC} and representational compactness~\cite{2017_light-weight-triangulation}) or physical (e.g., stress relief~\cite{2006_chenYong_mesh-based}). Meshing approaches vary; some directly triangulate the entire cell geometry using established algorithms~\cite{2017_light-weight-triangulation,2017_factedDiamondMapping,2017_prefab-cell_HGM,zou2024meta}, while others separately mesh nodes and trusses before assembling them into a coherent model~\cite{2005_HongqingWang_2Dsurface,2002_HongqingWang_conformal,2006_chenYong_mesh-based,2007_ChenYong_3DtextureMapping}. This modular approach proves advantageous in handling complex intersections and simplifying geometric calculations~\cite{2020_convexHull,2020_CHoCC}.

\textbf{Parametric surface-based methods.} 
Parametric surfaces offer compact representations compared to meshes, leveraging a small number of control points to describe complex geometries that would otherwise require thousands of triangles. Additionally, parametric surfaces are continuous and smooth. Given these advantages, van Sosin et al.~\cite{2021_spherePacking_conformal} employed parametric surfaces to represent microstructures. The cell geometry is constructed as an assembly of trimmed bi-quadratic rational B-spline surfaces, i.e., the node is a trimmed spherical surface, and half-beams are trimmed surfaces of revolution (Fig.~\ref{fig:surface}b). However, the process of trimming parametric surfaces (e.g., surface-surface intersections) is prone to errors, see Chapter 5 of~\cite{patrikalakis2002shape}. These methods are constrained by the requirement to adhere to rectangular topologies in surfaces being used, limiting their applicability to microstructures with intricate topologies.

\textbf{Subdivision surface-based methods.} 
Subdivision surfaces, which essentially are piecewise parametric surfaces defined over meshes of arbitrary topology, have both the topological flexibility of meshes and the smoothness of parametric surfaces~\cite{2006_study-on-subdivision-surface}. For this reason, subdivision surfaces have been used to model microstructures by designing specialized control meshes~\cite{2023_2Dsurface_Xiong_Subdivisional-modelling,2018_subdivision_2Dsurface,2019_differentialOffsetGrading}. The downside of such methods is that designing effective control meshes remains a non-trivial task, especially for cell geometries with intricate topology.

\textbf{Remarks:} 2D surface-based representations are evaluated schemes, where the shapes are described using explicit or evaluated geometric entities such as vertices, edges, and faces~\cite{requicha1982solid}. This makes them more convenient than other representations (e.g., implicit schemes) for downstream applications such as visualization and fabrication. However, they often demand higher memory usage and may exhibit reduced robustness compared to other representation methods. Current methods have primarily been validated on microstructures with simple geometries and at small scales. Further research is needed to enhance the practical utility of 2D surface-based representations, particularly for more complex and large-scale microstructures.

\subsubsection{3D volume-based representations}
\label{sec:3D}
3D volume-based representations encompass a variety of methods for modeling microstructures, each offering unique advantages and challenges. These methods include voxels~\cite{2007_3D-rep_Kou_Heterogeneous-modeling-review,2017_3D-rep_Aremu_voxel_trimmed-lattice}, polyhedrons~\cite{2013_3D-rep_Feng_FVM_macromolecules,2020_3D-rep_letov_bio-inspired_volumetric-cell,2020_topology-rep_Lei_Parametric-design-Voronoi-based,2022_polyhedral-voronoi-cancellous,2016_polyhedral-voronoi-scaffolds}, CSG~\cite{2019_3D-rep_Gupta_CSG_steady-lattice}, texture mapping~\cite{2007_ChenYong_3DtextureMapping,2017_factedDiamondMapping}, and volumetric splines~\cite{2023_3D-rep_Elber_trivariate-spline_review,2014_3D-rep_Wang_trivariate-spline_solid,2015_3D-rep_Lin_trivariate-spline_solid,2019_3D-rep_Dokken_trivariate-spline_solid,2017_3D-rep_Elber_trivariate-spline_Functional-Composition,2018_3D-rep_Massarwi_trivariate-spline_extension-Hierarchical,2021_3D-rep_Hong_trivariate-spline_Trimmed-Trivariate,2016_3D-rep_Massarwi_trivariate-spline_trimmed,2022_3D-rep_Gao_trivariate-spline_sample-to-porous,2023_Elber_implicit-conformal_CAD,2021_BsplineMapping,2023_3D-rep_THBspline_multilevel-porous}. Unlike 2D surface-based methods, these representations capture both microstructure boundaries and interiors, enabling precise characterization of heterogeneous properties. They are particularly compatible with topology optimization techniques such as the solid isotropic material with penalization (SIMP) method~\cite{2016_Liu_TO-SIMP-review}. They are also more robust in handling topology changes than the previous surface-based representations~\cite{2023_3D-rep_Elber_trivariate-spline_review}. For these reasons, many prefer to represent microstructures with volumetric methods (as detailed below) in spite of the memory overheads.

\begin{figure*}[t]
\centering
\includegraphics[width=0.9\textwidth]{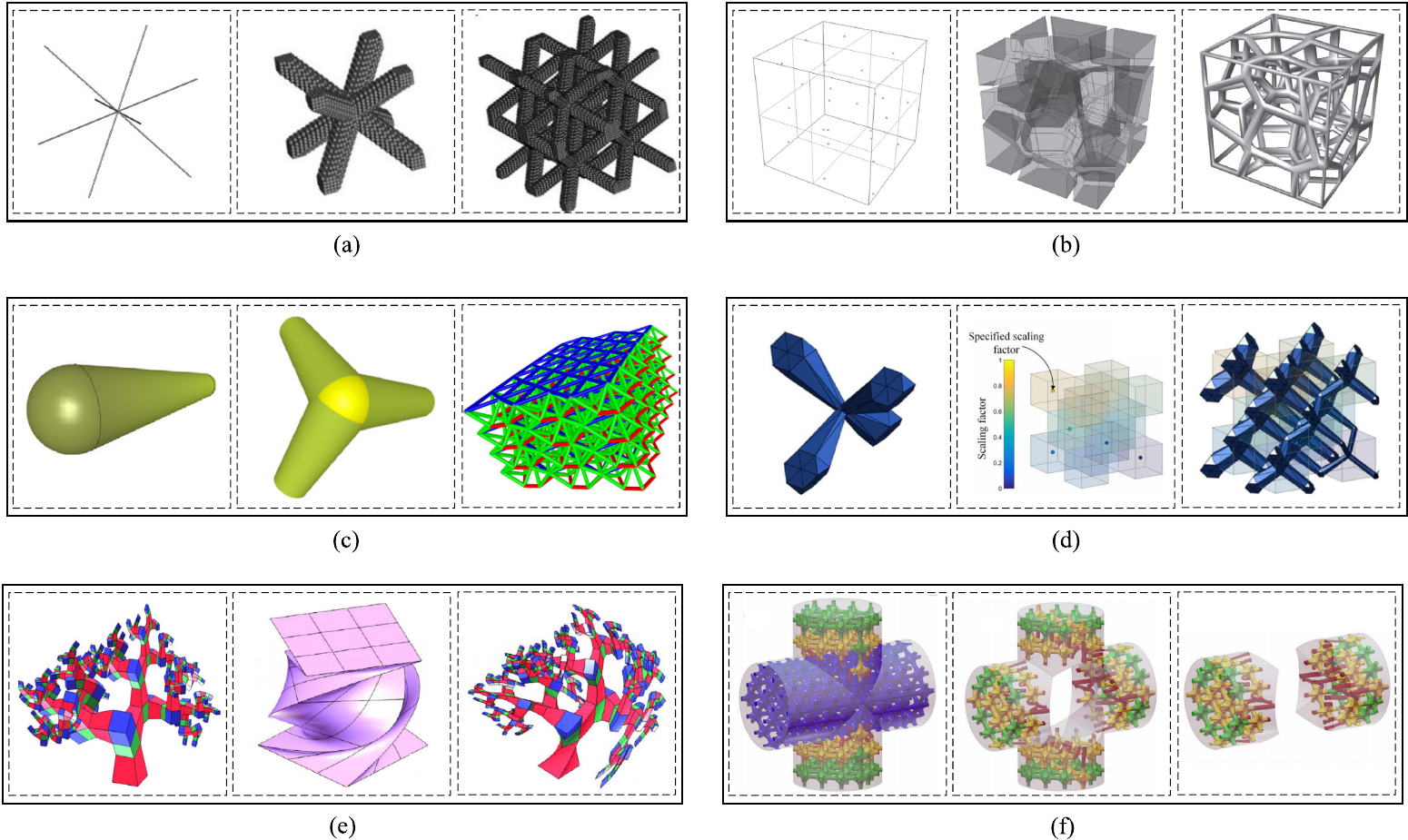}
\caption{Illustration of volume-based representations: (a) voxel-based lattice structure~\cite{2017_3D-rep_Aremu_voxel_trimmed-lattice}; (b) polyhedron mesh-based microstructure~\cite{2016_polyhedral-voronoi-scaffolds}; (c) CSG-based lattice structure~\cite{2019_3D-rep_Gupta_CSG_steady-lattice}; (d) 3D texturing-based microstructure~\cite{2017_factedDiamondMapping}; (e) spline-based microstructure~\cite{2018_3D-rep_Massarwi_trivariate-spline_extension-Hierarchical}; and (f) spline-based lattice structure generated in trimmed model~\cite{2021_3D-rep_Hong_trivariate-spline_Trimmed-Trivariate}.}
\label{fig:volume}
\end{figure*}

\textbf{Voxel-based methods.}
A voxel is a cube-like element in space, and a voxel model is a collection of voxels comprising a three-dimensional geometry
of interest~\cite{zou2023xvoxel}. Because all voxels in the collection are uniform, representing a porous structure is conceptually similar to representing a plain solid block, except for requiring more memory. As such, voxels have been used to represent intricate microstructures resulting from topology optimization, e.g.,~\cite{2022_Feng_TPMS-optimization-SIMP,2021_graded-TPMS-optimization_voxel-based,2017_Giga-voxel-structural-design_SIMP,2020_voxelFuse-lattice}. Furthermore, voxel model editing is very easy since all editing operations boil down to only two basic operations: voxel removal and voxel addition. Taking this as a basis, Aremu et al. ~\cite{2017_3D-rep_Aremu_voxel_trimmed-lattice} generated complex trimmed lattice structures by making use of voxel booleans on the macro boundary model and the cell geometry, as shown in Fig.~\ref{fig:volume}a.

\textbf{Polyhedral mesh-based methods.}
A polyhedron, consisting of polygons joined at their edges, is a generalized version of voxels; a collection of polyhedrons joined at their faces comprises a polyhedral mesh. This generalization offers better flexibility in representing 3D objects. Making use of this feature, Feng et al.~\cite{2013_3D-rep_Feng_FVM_macromolecules} proposed a multiresolution geometric representation method for biomolecular-like microstructures, and developed a set of microstructure visualization and analysis tools. Letov et al.~\cite{2020_3D-rep_letov_bio-inspired_volumetric-cell} proposed two bio-inspired heuristics to improve the generation and arrangement of polyhedrons, yielding more representational flexibility. However, this method's effectiveness was only demonstrated using simple cases that are 2D and have $<10$ cells.

\textbf{Voronoi-based methods.}
These methods can be considered an extension of polyhedral mesh-based methods, where polyhedrons are arranged following the pattern of 3D Voronoi di- agrams, as illustrated in Fig.~\ref{fig:volume}b. They are particularly effective for modeling microstructures with irregular topologies.  That is, different Voronoi diagrams can be produced by varying seed generation.  Seeds can be generated either randomly~\cite{2019_voronoi_random_lattice,2022_polyhedral-voronoi-cancellous,2023_voronoi_random,2018_polyhedral-voronoi-based-random} or based on specific constraints to tailor the microstructure’s properties~\cite{2016_polyhedral-voronoi-scaffolds,2018_voronoi_Controllable-Porous,2019_voronoi_control-network,2020_topology-rep_Lei_Parametric-design-Voronoi-based,2023_voronoi_control_stress-line,2024_voronoi_control-conformal-density,2024_stress-field-voronoi}, such as graded porosity. While Voronoi-based methods excel at creating highly irregular microstructures, it is important to note that generating 3D Voronoi diagrams with millions of seeds may be computationally demanding.

\textbf{CSG-based methods.}
CSG represents a solid as successive combinations (via regularized Boolean operations) of primitive shapes~\cite{1980_Requicha_CSG}. Gupta et al.~\cite {2019_3D-rep_Gupta_CSG_steady-lattice} used this way to represent (as well as construct) lattice structures as unions of solid balls and solid cone sections, as depicted in Fig.~\ref{fig:volume}c. This method is computationally simple, but it is designated for the special microstructure type considered in the paper. For more complex, general microstructures, further development is needed.

\textbf{3D Texturing-based methods.} 
Leveraging techniques from computer graphics, texturing maps 2D images onto 3D objects. Similarly, by constructing a mapping between a microstructure's cell geometry and a pre-defined 3D texture, Chen et al.~\cite{2006_chenYong_mesh-based,2007_ChenYong_3DtextureMapping,2018_2Dsurface_chen_mapping-deformation} and Peng et al.~\cite{2004_Peng_volume-texturing} developed a series of texturing-like microstructure modeling methods that can handle intricate cell geometries. Yoo~\cite{2012_Yoo_voxelMapping,2011_Yoo_hexMapping} utilized a similar approach for TPMS-based porous scaffold design. Dumas et al.~\cite{2017_factedDiamondMapping} extended this idea to include a new type of cell geometry, i.e., the faceted diamond cell consisting of truncated tetrahedral nodes and four hexagonal struts, as shown in Fig.~\ref{fig:volume}d. It should, however, be noted that unexpected distortions within a cell and discontinuities between adjacent cells may happen when using texturing-based methods.

\textbf{Volumetric spline-based methods.}
Like the texturing-based methods, volumetric spline-based methods~\cite{2014_3D-rep_Wang_trivariate-spline_solid,2015_3D-rep_Lin_trivariate-spline_solid,2019_3D-rep_Dokken_trivariate-spline_solid,2017_3D-rep_Elber_trivariate-spline_Functional-Composition} are also mapping-based. The basic idea is to use a trivariate tensor-product spline volume to map from a canonical 3D parametric domain to the 3D design space. Then by tiling the parametric domain and filling each tile with some simple micro-level structures, the spline volume can deform these structures into a more complex form residing in the design space, as shown in Figs.~\ref{fig:volume}e, and~\ref{fig:volume}f. In addition, the spline volume's control points offer an intuitive way of designing and editing the mapping, and therefore the final microstructure. The development of volumetric spline-based methods has two lines of research. The first line is concerned with parametric domain tiling~\cite{2022_3D-rep_Gao_trivariate-spline_sample-to-porous,2023_Elber_implicit-conformal_CAD}, for example, extending from the common beam-based micro-level structures to hierarchical ones~\cite{2018_3D-rep_Massarwi_trivariate-spline_extension-Hierarchical}. The second line focuses on the design of spline volumes~\cite{2021_BsplineMapping,2023_3D-rep_THBspline_multilevel-porous}, for example, through function composition~\cite{2017_3D-rep_Elber_trivariate-spline_Functional-Composition} or control point optimization~\cite{2019_3D-rep_Antolin_trivariate-spline_optimize-tile}, see Elber's recent review~\cite{2023_3D-rep_Elber_trivariate-spline_review} for more details. Volumetric spline-based methods are promising, but there is one practical issue that needs to be solved: tensor-product B-spline volumes must be a cuboid in topology, leading to limited applicability. Trimming may help but cannot solve all the problems together~\cite{2021_3D-rep_Hong_trivariate-spline_Trimmed-Trivariate,2016_3D-rep_Massarwi_trivariate-spline_trimmed}. For example, it may have computational robustness issues, as well as microstructure discontinuity issues between trimmed volumes~\cite{2023_3D-rep_Elber_trivariate-spline_review}.

\textbf{Remarks:} 3D volume-based representations have the advantage of the ability to characterize heterogeneous geometric/physical properties and their well-alignment with topology optimization methods. However, their memory consumption is often high, and their applicability to CS/HCS is yet to be demonstrated.

\subsubsection{3D implicits-based representations}
\label{sec:4D}
Compared to 3D volume-based representations, 3D implicits-based representations introduce an additional scalar function. Methods of this kind are known as implicit representations and include signed distance fields, implicit B-splines, and TPMS, among others. Central to all these approaches is a scalar function defined across the 3D space, with its isocontour used to indicate the boundary between solid and void in microstructures. Thus, designing a microstructure becomes the task of manipulating this scalar function, which can be easily done by existing algorithms such as F-rep-based Booleans and blending~\cite{groen2021multi}. Consequently, implicits have found extensive applications in microstructure design and manufacturing, and those methods deserve a review paper by themselves, e.g.,~\cite{2018_Li_implicit-Modeling-AM-review,2022_Feng_TPMS-review}; there have also been successful commercial software packages made available, e.g., nTopology~\cite{nTopology} and IceSL~\cite{iceSL}. Here, we provide a sampling of those methods, focusing on microstructure representation rather than design and optimization methods.

\begin{figure*}[t]
\centering
\includegraphics[width=0.94\textwidth]{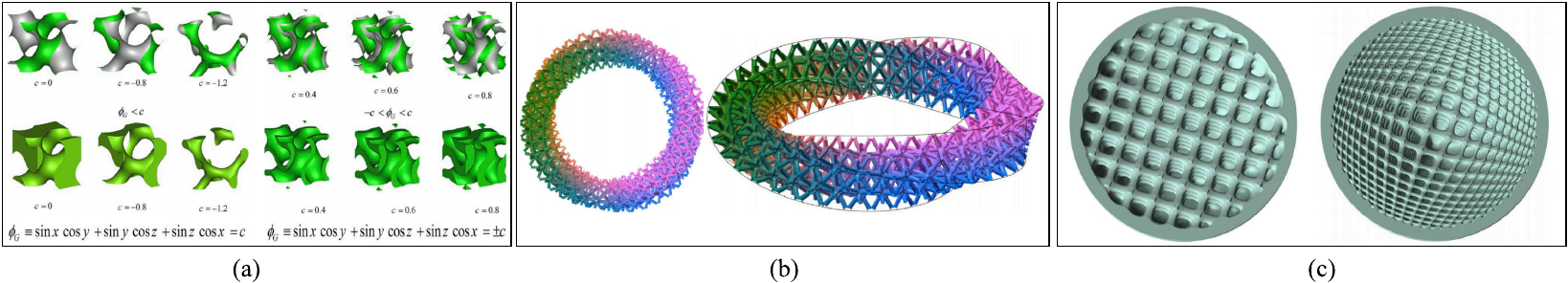}
\caption{Illustrations of implicits-based representations: (a) TPMS shape control with trigonometric equations~\cite{2020_4Dscale_network-sheet-TPMS}; (b) free-form shape control using spline functions~\cite{2023_Elber_implicit-conformal_CAD}; and (c) general shape control based on F-rep~\cite{2011_Pasko_FrepLattice}.}
\label{fig:scale}
\end{figure*}

Depending on the control parameters a representation scheme gives to the user, implicits-based representation schemes may be classified into three categories: specialized, free-form, and general. Specialized representations admit a very special class of equations to define the scalar function, thereby leaving a very limited number of parameters for tuning microstructure shape. One typical example of such representations is TPMS (Fig.~\ref{fig:scale}a). It is defined solely by trigonometric functions, and only the cell geometry thickness and periodicity are editable~\cite{2022_4D-rep_Wang_TPMS}. Another example is the method proposed by Gupta et al.~\cite{2019_3D-rep_Gupta_CSG_steady-lattice}, where lattices are represented implicitly by a parameterized Lattice Maker Applet (LMA).
Free-form representations relax the defining equations to the family of spline functions, where control points offer greater flexibility in shaping microstructures (Fig.~\ref{fig:scale}b)~\cite{2023_Elber_implicit-conformal_CAD,2018_3D-rep_Massarwi_trivariate-spline_extension-Hierarchical}. 
The third category of general representations adopts general implicit modeling frameworks (e.g., F-rep framework~\cite{1995_Pasko_F-rep}) to attain arbitrary control over microstructure shape~\cite{2011_Pasko_FrepLattice,2013_4D-rep_Fryazinov_Frep_multi-scale,2020_subdivision-contour-extraction,2024_F-rep_zhao,2020_spinodoid-metamaterials,2020_level-surface-Guassian-kernel}, see Fig.~\ref{fig:scale}c for an example. Loosely, F-rep is a notion that can be used interchangeably with the implicit representation, and it makes no particular assumption on the defining equations---any continuous real-valued function is allowed, and thus the design space is unlimited.

\textbf{Remarks:} 3D implicits-based representations can handle a very wide variety of microstructures. The associated modeling algorithms are also quite robust. However, interactively controlling and visualizing these representations is not trivial~\cite{1995_Pasko_F-rep}. Also, when using this kind of representation, many microstructures cannot be represented analytically, and we need to resort to discrete scalar fields, resulting in a very high memory consumption. Hence, 3D implicits-based representations are most suitable for analytically describable microstructures or those on a small scale.

\subsubsection{Hybrid representations}
\label{sec:hybrid}
As the name suggests, hybrid representations combine multiple representations mentioned above to get the best of both worlds. For example, we can combine the volumetric spline-based method~\cite{2023_3D-rep_Elber_trivariate-spline_review} with the 3D implicits-based representation~\cite{1995_Pasko_F-rep} to offer more flexibility in controlling cell geometry~\cite{2023_Elber_implicit-conformal_CAD}. 
However, despite their obvious advantages, hybrid representations are surprisingly underexplored in the literature. Only a few combinations have been reported. Wang et al.~\cite{2005_HongqingWang_2Dsurface} integrated B-rep solid modeling and polygonal representation to improve the efficiency of lattice structure evaluation. Similarly, Vongbunyong et al.~\cite{2017_prefab-cell_HGM} presented a hybrid B-Rep and polygon method for the rapid generation of fabrication-ready cell geometry. Gupta et al.~\cite{2019_hybrid_Gupta_CSG+CST+Brep} combined traditional CSG, B-rep, and constructive solid trimming to attain a good trade-off among representational exactness, memory efficiency, and geometric query efficiency. 

\textbf{Remarks:} Hybridizing different representation methods offers the potential for enhancing computational efficiency, robustness, and modeling consistency simultaneously. However, research in this area is limited, and significant challenges remain, including (1) the need for automatic conversion mechanisms between constituent representations to ensure consistency and (2) the development of memory-efficient data structures, particularly critical for CS/HCS.

\def\tabularxcolumn#1{m{#1}}
\begin{table*}[b]
\caption{Typical design-oriented modeling algorithms of microstructures.}
\centering
\setlength\extrarowheight{8pt}
\begin{tabularx}{\textwidth}{
    >{\centering\arraybackslash}>{\hsize=.7\hsize\linewidth=\hsize}X
    >{\centering\arraybackslash}>{\hsize=.8\hsize\linewidth=\hsize}X
    >{\centering\arraybackslash}>{\hsize=1.5\hsize\linewidth=\hsize}X
   }
    \Xhline{1.2pt}
     Types&Operations &Papers\\
    \Xhline{1.2pt}
    \multirow{6}{*}{Geometrical Operations} & Querying & Queries for point membership classification, minimum distance, ray intersection~\cite{2019_hybrid_Gupta_CSG+CST+Brep}; accelerated ball-interference query against steady lattices~\cite{2019_Range-finder_query}\\

    & Boundary evaluation & Approximation~\cite{2020_CHoCC}; triangulation~\cite{2017_light-weight-triangulation,zou2024meta}; evaluation robustness~\cite{2020_convexHull}; TPMS-to-Spline conversion~\cite{zou2024tpms2step}\\
    
    & Blending &Point-based blending~\cite{2006_chenYong_mesh-based}; R-function-based blending~\cite{2011_Pasko_FrepLattice}; analytical blending~\cite{2020_quador_fillets}; convolution surface-based blending~\cite{2021_1D-skeleton_Zouqiang_convolution-surface}; TPMS composition-based blending~\cite{2024_blending_TPMS-based-fillet}\\
    & Offsetting & Offsetting for F-rep-based lattices~\cite{2011_Pasko_FrepLattice}\\
    & Simplifying & Simplify cell geometries by replacing spherical nodes with convex hulls~\cite{2020_convexHull,2020_CHoCC,2005_1D_convex-hull_solidifying-wireframe}\\
    & Conforming & Conformal design of microstructures based on global parameterization~\cite{2017_3D-rep_Elber_trivariate-spline_Functional-Composition,2021_BsplineMapping,2018_TsplineMapping,2023_Elber_implicit-conformal_CAD,2002_HongqingWang_conformal,2019_3D-rep_Antolin_trivariate-spline_optimize-tile}, mapping cells to volumetric meshes~\cite{2012_Yoo_voxelMapping,2011_Yoo_hexMapping,2018_2Dsurface_chen_mapping-deformation,2024_conformal_TPMS+lattice}, constructing boundary skins~\cite{2017_3D-rep_Aremu_voxel_trimmed-lattice}\\

    \hline
    \multirow{2}{*}{Topological Operations} & Booleans & Trimming microstructures~\cite{2017_3D-rep_Aremu_voxel_trimmed-lattice}; merging multiple microstructures~\cite{2024_multiple-sized_TPMS_merging}\\
    & Hollowing & Hollowing via Voronoi diagram of ellipses~\cite{2018_support_Lee_self-support-infill_ellipse-hollowing}; shape optimization-based hollowing~\cite{2018_design-operation_Support-Free-Hollowing,2014_design-operation-build-to-last-hollowing}\\
    \Xhline{1.2pt}
\end{tabularx}%
\label{tab:statistics-of-design-operation}%
\end{table*}%

\section{Microstructure modeling algorithms}
\label{sec:operation}
Microstructure representations are mathematical and/or computational models for the description of microstructure shapes. Microstructure operations, on the other hand, are algorithms used to create, edit, and analyze those models to support simulation, optimization, fabrication, etc. This paper classifies those algorithms into two major categories: design-oriented (Sect.~\ref{sec:design-operation}) and manufacturing-oriented (Sect.~\ref{sec:manufacturing-operation}). The former is further divided into two groups: geometric and topological. Refer to Table~\ref{tab:statistics-of-design-operation} for details. The latter category focuses on four fundamental AM processes: part orienting, slicing, support generation, and path infilling.  Refer to Table~\ref{tab:manufacturing-operation} for a snapshot of them.

\subsection{Design-oriented operations}
\label{sec:design-operation}
Design-oriented operations are used primarily in the creation, modification, and analysis of microstructures. In traditional CAD, many such operations have been developed and proved effective. Unfortunately, only a few of them have been adapted to microstructure modeling, including Boolean, hollowing, blending, offsetting, simplifying, boundary evaluation, and querying. This work groups them into the geometric and topological categories, depending on whether there are significant topology changes and/or shape variations, especially those at the meso- or macro-level, during the operation.

\subsubsection{Geometrical operations}
\textbf{Querying.} Geometric queries, such as point membership classification, distance query, and integral properties, are important to microstructure modeling. Given that microstructures have a large number of elements, those queries are time-consuming. To address this problem, Kurzeja et al.~\cite{2019_Range-finder_query} presented a RangeFinder algorithm to accelerate ball-interference queries for regular/semi-regular lattice structures. Gupta et al.~\cite{2019_hybrid_Gupta_CSG+CST+Brep} proposed a method to compute integral properties by decomposing lattice structures into disjoint parts, addressing efficiency and accuracy issues. Overall, despite the obvious importance of querying operations, their research and engineering remain underexplored.

\textbf{Boundary evaluation.} Boundary evaluation operations take as input an unevaluated description (e.g., parametric and implicit) of a microstructure and output its boundary representation to be used in downstream applications like data exchange, visualization, analysis, manufacturing, etc. Different from conventional solid modeling, boundary evaluation of microstructures needs to handle a much larger number of geometric elements and a much more complex way of how elements intersect with each other. Therefore, evaluation efficiency and robustness are two problems requiring special attention. To attain high efficiency, Wu et al.~\cite{2020_CHoCC} developed a simple, approximate method to evaluate the nodal geometry of lattice structures. Chougrani et al.~\cite{2017_light-weight-triangulation} provided an efficient triangulation method to generate lightweight boundary meshes for lattice structures. Recent advancements, such as those by Zou et al.~\cite{zou2024meta}, leverage GPU techniques to extend boundary evaluation to billion-scale lattice structures. Verma et al.~\cite{2020_convexHull} improve evaluation robustness by simplifying element connections with the help of convex hulls, at the price of modeling errors though. Besides lattice structures, there are also boundary evaluation methods developed for TPMS structures; for example, the method presented by Zhao et al.~\cite{zou2024tpms2step} can convert implicit TPMS models to $C^2$ continuous B-spline surfaces.

\textbf{Blending.} Blending operations generate smooth transitions between two or several geometries. They are typically used in CAD to create fillets to ease sharp features and relieve stresses. In microstructure modeling, the relevant research may be initialized by Chen~\cite{2006_chenYong_mesh-based}, where a point-based method was utilized to add blends to lattice structures (represented in the mesh format). Later, Pasko et al.~\cite{2011_Pasko_FrepLattice} presented a more efficient and robust microstructure blending method by making use of F-rep, where adding blends are converted to simple function additions~\cite{pasko2005bounded}. Liu et al.~\cite{2021_1D-skeleton_Zouqiang_convolution-surface} extended this approach using convolution surfaces to attain more natural and robust blending. Gupta et al.~\cite{2018_quador} and Cirak and Sabin~\cite{2020_quador_fillets} provided analytical methods for adding blends to lattice structures. Iandiorio et al.~\cite{2024_blending_TPMS-based-fillet} replaced cylindrical trusses with TPMS-based trusses, which have natural blends. 

\textbf{Offsetting.} Offsetting operations generate expanded or shrunk versions of geometric objects. They are useful for tolerance analysis, clearance tests, tool path generation, etc. In microstructure modeling, offsetting operations are important to attain physical property variations. Because a microstructure's cell geometry could be very complex, conventional explicit offsetting methods~\cite{rossignac1986offsetting} may have robustness issues, and distance field-based implicit methods are often the method of choice~\cite{2011_Pasko_FrepLattice}.


\textbf{Simplifying.} Simplifying operations reduce the complexity of a geometric object while keeping the overall shape, volume, and/or boundaries as much as possible. In microstructure modeling, simplification is often done at the cell geometry level and used to ease stress concentration and improve computational robustness. For example, spherical nodes in a lattice structure are replaced with convex hulls to avoid sharp edges and complex surface-surface intersections~\cite{2020_convexHull}. Similar ideas were also presented in~\cite{2020_CHoCC,2020_1D-skeleton_Liang_VDF-conformal,2005_1D_convex-hull_solidifying-wireframe}. It should be noted that little attention has been paid to simplification at the macro- or meso-level, which otherwise could offer huge potential for visualization and simulation.

\textbf{Conforming.} Conforming operations are used to generate boundary-conformed microstructures. Different strategies have been developed to implement this operation, depending on the microstructure type considered. One strategy is first constructing a global parameterization (using volumetric splines) of the macro-level shape and then deforming a regular microstructure into a shape-conformed microstructure through the parameterization~\cite{2021_BsplineMapping,2018_TsplineMapping,2023_Elber_implicit-conformal_CAD,2017_3D-rep_Elber_trivariate-spline_Functional-Composition,2019_3D-rep_Antolin_trivariate-spline_optimize-tile}. With the help of parameterization, we can also generate shape-conformed lattice structures directly in the design space by connecting nodes on adjacent iso-parametric surfaces~\cite{2002_HongqingWang_conformal}. As such, cell geometries closely follow the iso-parametric surfaces, and therefore the overall shape (i.e., the outmost iso-parametric surface). Another conforming strategy is combining the cell decomposition method with the cell filling method~\cite{2012_Yoo_voxelMapping,2011_Yoo_hexMapping,2018_2Dsurface_chen_mapping-deformation,2024_conformal_TPMS+lattice}. Basically, it first generates a hex-mesh for the design space of interest and then fills each mesh element with a deformed version of pre-defined cell geometry. Note that both parameterization and hex-meshing remain open problems by themselves. In addition to the above two strategies, shape conformity of microstructures can also be achieved by using the signed distance field (SDF) of the boundary shape. For example, Liu et al.~\cite{2024_voronoi_control-conformal-density} and Fryazinov et al.~\cite{2015_interior-SDF_conformal} constructed conformal Voronoi diagrams within the boundary shape represented by the SDF, then generated microstructures based on such Voronoi diagrams. Another relevant study by Lee et al.~\cite{2020_SDF-conformal-CHequation} integrated the SDF with the nonlocal Cahn–Hilliard (CH) equation and generated conformal porous structures by solving this CH equation. Note that the computation and manipulation of distance fields could be computationally intensive. Besides the above strategies, a brute-force strategy has also been reported~\cite{2017_3D-rep_Aremu_voxel_trimmed-lattice}. It first trims a regular microstructure against the intended macro-level shape and then thickens the macro-level shape's boundary to such a level that any dangling pieces in the trimmed microstructure are to be covered. This method is computationally simple, but the resulting microstructures are not optimized.

\subsubsection{Topological operations}
\textbf{Booleans.} Boolean operations are a powerful tool for creating complex geometries through adding, subtracting, or intersecting simple geometries. In microstructure modeling, Boolean operations are often used for trimming microstructures against macro-level shape~\cite{2017_3D-rep_Aremu_voxel_trimmed-lattice} or merging multiple microstructures~\cite{2024_multiple-sized_TPMS_merging,2023_3D-rep_Elber_trivariate-spline_review}. Currently, Boolean operations on microstructures are implemented by resorting to voxel modeling and/or implicit modeling methods that have simple and robust algorithms made available. A typical example is the software nTopology~\cite{nTopology}. However, both voxels and implicits are compute-intensive, necessitating further developments in microstructure-specific Boolean operations, especially when it comes to SS and HCS modeling.

\textbf{Hollowing.} Hollowing operations generate micro-holes within geometries. They can be viewed as a specialized version of Boolean operations, but augmented with hole distribution design methods. For instance, Voronoi diagrams have been used to optimize hole distribution~\cite{2018_support_Lee_self-support-infill_ellipse-hollowing,2014_design-operation-build-to-last-hollowing}. In the method presented by Wang et al.~\cite{2018_design-operation_Support-Free-Hollowing}, offsetting operations are used to optimize the holes to attain self-supporting features.


\def\tabularxcolumn#1{m{#1}}
\begin{table*}[!b]
\caption{Typical manufacturing-oriented modeling algorithms of microstructures.}
\centering
\setlength\extrarowheight{8pt}
\begin{tabularx}{\textwidth}{
    >{\centering\arraybackslash}>{\hsize=.6\hsize\linewidth=\hsize}X
    >{\centering\arraybackslash}>{\hsize=.9\hsize\linewidth=\hsize}X
    >{\centering\arraybackslash}>{\hsize=1.5\hsize\linewidth=\hsize}X
   }
    \Xhline{1.2pt}
     Operations & Functionalities &References\\
    \Xhline{1.2pt}
    \multirow{2}{*}{Part Orienting} & Single-orientation printing & Part orientation optimization considering manufacturability, mechanical performance or surface quality~\cite{2020_orientation_multicriteria_processability,2023_orientation_fatigue-of-lattice}; part orientation optimization with accelerated computation speed~\cite{2020_orientation_optimization-with-first-order-deriviation} \\
    & Multi-orientation printing & Multi-orientation printing of lattice structures~\cite{2021_slicing_multi-directional}\\
    \hline
    \multirow{2}{*}{Layer Slicing} & Efficiency improvement & Culling-based slicing~\cite{2021_slicing_efficiency_matrix-oriented-data-structure,2021_slicing_efficiency_Slice-Traversal}; out-of-core slicing~\cite{2021_1D-skeleton_Zouqiang_convolution-surface}; ray tracer-based slicing~\cite{2021_Maltsev_slicing_acceleration}\\
    & Robustness enhancement & F-rep-based slicing~\cite{2019_slicing_TPMS_CAD,2021_1D-skeleton_Zouqiang_convolution-surface,2024_slicing_microcellular-structures}; persistent homology-based slicing~\cite{2022_PersistentHomology,2023_slicing_persisitentHomology_adapative}; ray tracer-based slicing~\cite{2021_slicing_efficiency_matrix-oriented-data-structure} \\
    \hline
    \multirow{2}{*}{Support Generation} & Posterior methods &Self-supported unit cell-based design~\cite{2019_support_Hu_self-support-design_TPMS,2016_support_Aremu_self-support-design_BCC,2021_support_Zhou_self-support-design_multi-fold-rotational-symmetry}; multi-axis printing~\cite{2021_slicing_multi-directional}; optimal orientation~\cite{2020_orientation_multicriteria_processability}; column-like support structure~\cite{2022_support_column-for-lattice}; tree-like support structure~\cite{2021_support_Wang_minimize-geometry_Support-point-determination,2023_support_grid-tree-for-lattice} \\
    & Priori methods &Topology/geometric optimization-based support generation~\cite{2023_support_Wang_self-support-deformation_subdivision-simplicification,2016_support_Wu_self-support-infill_rhombic,2013_support_Wang_minimize_cost-effective,2018_support_Lee_self-support-infill_ellipse-hollowing,2017_support_Wang_self-support-infill_support-free-frame,2021_support_Liu_self-support-infill_TO-shell} \\
    \hline
    \multirow{5}{*}{Path Infilling} & Direction-parallel path infilling & Solid-based microstructures~\cite{2018_slicing_F-rep}; shell-based microstructures~\cite{ding2021stl,2024_direction-or-contour-path-TPMS}, beam-based microstructures~\cite{2024_infill_direction-lattice}\\
    & Contour-parallel path infilling & TPMS microstructures~\cite{2024_direction-or-contour-path-TPMS}, lattice microstructures~\cite{2024_infill_truss-spherical-path-planning}, honeycomb-like microstructures~\cite{2019_infill_contour-complex-geometries,2023_infill_contour-continuous-honeycomb}.\\
    & Space-filling path infilling & Solid-based microstructures~\cite{2016_infill_Fermat-Spirals}\\
    &Single-stroke path infilling & Solid-based microstructures~\cite{2023_infill_honeycome-infill,2020_infill_sparse-infill,2019_infill_Porous-Structures,2022_contimuous_CGF} \\
    &Hybrid path infilling & Combination of zigzag and contour paths~\cite{2022_Bi_infill_Continuous-contour-zigzag, 2022_Xia_infill_Continuous-contour-zigzag,2022_hybrid-path_zigzag-contour}; combination of zigzag and spiral path~\cite{2014_infill_hybrid_zigzag-spiral}\\
    
    \Xhline{1.2pt}
\end{tabularx}%
\label{tab:manufacturing-operation}%
\end{table*}%

\subsection{Manufacturing-oriented operations}
\label{sec:manufacturing-operation}
\begin{figure}[htbp]
\centering
\includegraphics[width=0.3\textwidth]{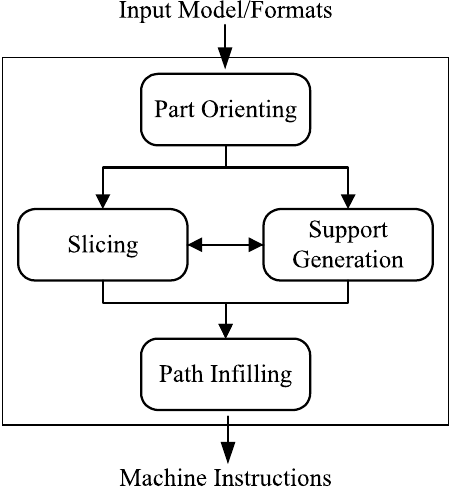}
\caption{Microstructure process planning pipeline.}
\label{fig:printing_pipeline}
\end{figure}

Geometric operations in this category are used to plan the manufacturing process of microstructures. As depicted in Fig.~\ref{fig:printing_pipeline}, these operations primarily encompass part orienting (Sect.~\ref{sec:part_orienting}), slicing (Sect.~\ref{sec:layer_slicing}), support generation (Sect.~\ref{sec:support}), and path infilling (Sect.~\ref{sec:path_infilling}). Actually, most of these operations are just customized versions of general-purpose ones used in the conventional additive manufacturing pipeline. For this reason, the following review will first outline the foundational strategies of these general-purpose methods before going into their specialized adaptations for the unique requirements of microstructure manufacturing, refer to Table~\ref{tab:manufacturing-operation} for a sampling of such studies. 

\subsubsection{Part orienting}
\label{sec:part_orienting}
Part orienting determines how the part to be printed is oriented relative to the building direction. This decision significantly impacts surface quality, printing time, support structure volume, part strength, and packing efficiency. The literature distinguishes between two main strategies: single-orientation printing and multi-orientation printing. Single-orientation printing sticks to a single building direction during the entire printing process, while multi-orientation printing allows varying building directions.

\textbf{Single-orientation printing.} In single-orientation printing, the part is built using a fixed orientation throughout the printing process. Orientation determination typically involves formulating an energy-minimization problem, often with constraints related to manufacturability. Optimization objectives include improving surface quality~\cite{2021_orientation_dimensional-accuracy,2021_orientation_local-region-curvature_low-volume-error}, enhancing mechanical properties~\cite{2020_orientation_anisotropic-behavior,2018_orientation_tensile-strength}, reducing printing time~\cite{2022_orientation_multi-objective_time-and-supports,2017_orientation_build-time}, and minimizing support structures~\cite{2020_orientation_single-objective_supports,2016_orientation_single-objective_over-hang}. Recent advancements include the application of machine learning methods to optimize orientation based on historical data and models~\cite{2023_orientation_AI-based,2023_single-orientation_Reinforcement-Learning,2023_part-orientation_CNN}, but there is still room for improvement, especially for datasets and machine learning models. For example, convolutional neural networks (CNNs) have been employed for part classification and orientation prediction in~\cite{2023_orientation_AI-based,2023_part-orientation_CNN}, but CNNs are primarily designed for image data, while many 3D prints are represented in formats like STL or B-rep, which do not assume the regular structure found in images. Consequently, there is a need for more specialized machine learning models tailored to 3D prints. Moreover, the datasets used in these studies were either limited in size~\cite{2023_single-orientation_Reinforcement-Learning,2023_part-orientation_CNN} (with fewer than 10,000 3D models) or narrow in scope~\cite{2023_orientation_AI-based} (focusing mainly on feature-based mechanical components). To develop robust AI-based methods, sufficiently large and diverse datasets that encompass various design patterns are essential.

For microstructures, orientation determination follows the same strategy as stated above, e.g.,~\cite{2023_orientation_fatigue-of-lattice,2020_orientation_multicriteria_processability,2020_orientation_optimization-with-first-order-deriviation}. However, two special challenges should be noted. First, compared to traditional parts, microstructures may have many discrete overhang pieces, leading to more complex manufacturability constraints. Also, optimization efficiency is a big concern. To solve these challenges, Nguyen et al.~\cite{2020_orientation_optimization-with-first-order-deriviation}  accelerate the optimization by introducing an efficient convex approximation-based algorithm.

\textbf{Multi-orientation printing.} There are applications where a part consists of several components, each of which has its own optimal building direction. To handle such cases, two main strategies have been developed. 
The first strategy combines multiple optimized directions from individual components into a comprehensive building direction for the entire part. For instance, Zhang et al.~\cite{2017_support_Zhang_optimal-orientation_multi-part} and Abdulhameed et al.~\cite{2022_orientation_multi-part_combined-direction} used a feature-based method to generate candidate orientations for each part, and then applied the genetic algorithm to search the optimal combination of part build orientations. However, this way of working can only achieve sub-optimal solutions.
The other strategy is built upon multi-axis printing techniques and allows each component to be built in its optimal orientation. Guo et al.~\cite{2023_orientation_Multi-orientation_model-segmentation} first segmented the model using a Reeb graph, then the optimal build direction of each sub-model was found by minimizing the volumetric error. Alternatively, Wu et al.~\cite{2020_learning-multi-directional} employed a machine learning-based approach to find the optimal sequence of the clipping planes. A similar method has also been presented by Huang et al.~\cite{2016_FrameFab}. If we further relax the restrictions to allow building directions to vary continuously, any point of the part can have its own building direction. Then multi-axis support-free printing becomes possible, which is particularly useful for printing microstructures~\cite{2021_slicing_multi-directional}.

\subsubsection{Slicing}
\label{sec:layer_slicing}
The slicing operation cuts a part to be printed into a sequence of horizontal layers to facilitate support structure and infill path calculations in additive manufacturing processes. (Curved layers are also possible due to the emerging multi-axis 3D printers~\cite{2018_support_Dai_self-support-multi-axis_TOG,2021_slicing_non-planar_mathematical-functions,2021_slicing_non-planar_isothermal-surface,2023_slicing_non-planar_isothermal-surface,2021_slicing_multi-directional,2023_slicing_non-planar_hybrid,2022_support_Zhang_self-support_multi-axis}.) Two major calculations are involved in slicing: determination of each slicing plane's height and Boolean operations between each slicing plane and the part model. For the former, two general strategies have been developed: uniform and adaptive. The uniform strategy has a fixed layer thickness, which is easy to implement but results in low surface quality (e.g., the stairstep effect). The adaptive strategy, which determines the layer thickness based on a given precision limit, can improve surface quality, but their calculations are more complex~\cite{1998_localAdaptiveSlicing,1994_adaptive_slicing_cupsHeight,2002_adaptive_slicing_surfaceRoughness,2000_adaptive_slicing_areaDeviation,2005_adaptive_slicing_volumeDeviation,2019_adaptive_slicing_metricProfile,2022_slicing_adaptive_relative-volume-error}.

For the latter Boolean operations, the main focus is robustness towards intersecting the slicing plane and the part model. Traditionally, the intersection is carried out directly between the slicing plane and the part's boundary model, including parametric surface models~\cite{2017_slicing_Bspline-volume,2004_slicing_NURBS_CAD,2018_slicing_T-spline} and triangular mesh models~\cite{2015_stl_improved_contour_construction,2017_stl_optimal_slicing,2018_stl_real-time_slicing,2021_stl_efficient_slicing}. However, slicing boundary models is notoriously known to be unstable, a consequence of many singular cases. To solve this issue, implicit model-based methods have been developed~\cite{2013_slicing_intersection-free_topological-faithful,2020_subdivision-contour-extraction,2013_slicing_LDNI,2021_1D-skeleton_Zouqiang_convolution-surface}. They first convert the part model into an implicit representation, e.g., distance fields~\cite{2013_slicing_intersection-free_topological-faithful} or convolutional fields~\cite{2021_1D-skeleton_Zouqiang_convolution-surface}, and then use the point membership classification operator to classify points on the slicing plane into three categories of IN-Solid, ON-Boundary, and OUT-Solid (which can be readily used to generate infill paths and support structures). For implicit models, point membership classification is easy and reliable. 

When it comes to microstructures, most of the above methods are applicable, but the slicing problem here is more efficiency- and robustness-demanding because of the higher geometric complexity and the data exploration problem. To this end, several customizations/improvements have been made, as follows.

\begin{figure*}[t]
    \centering
    \includegraphics[width=0.9\textwidth]{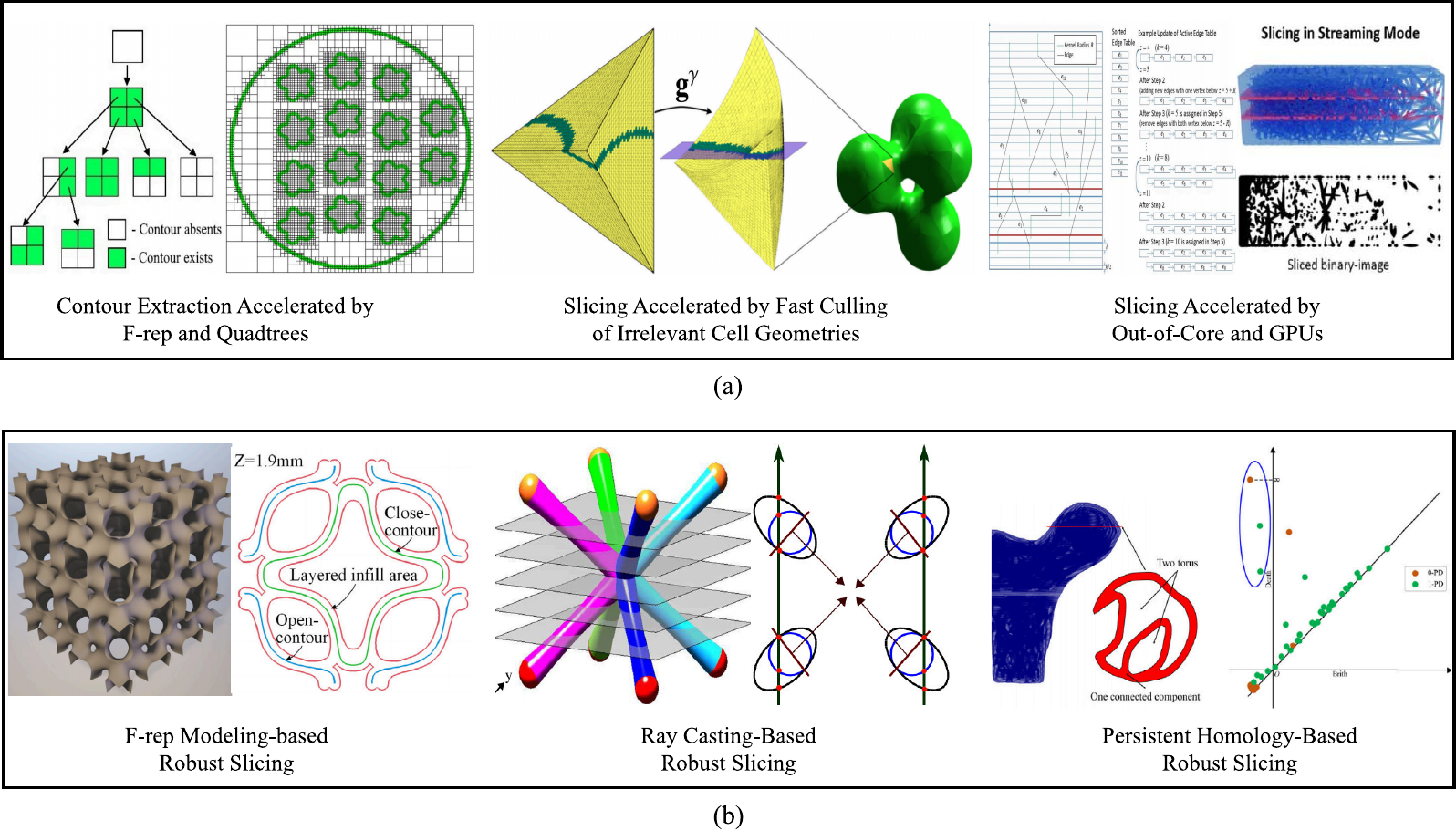}
    \caption{Examples of microstructure slicing methods: (a) slicing accelerated by quadtrees, fast querying, and out-of-core methods~\cite{2021_Maltsev_slicing_acceleration,2021_slicing_efficiency_Slice-Traversal,2021_1D-skeleton_Zouqiang_convolution-surface}; and (b) robustness enhanced by F-rep, ray tracing, and persistent homology methods~\cite{2019_slicing_TPMS_CAD, 2021_slicing_efficiency_matrix-oriented-data-structure,2023_slicing_persisitentHomology_adapative}.}
    \label{fig:slicing_results}
\end{figure*}

\textbf{Efficiency Enhancement.} Maltsev et al.~\cite{2021_Maltsev_slicing_acceleration} employed a combination of quadtrees, KD-trees, and R-trees to accelerate the slicing process. Mustafa et al.~\cite{2021_slicing_efficiency_matrix-oriented-data-structure} designed a matrix-based data structure to accelerate lattice structure topology querying and to facilitate multi-threaded slicing. Youngquist et al.~\cite{2021_slicing_efficiency_Slice-Traversal} presented a fast querying method for culling cell geometries irrelevant to the current slicing plane, thereby accelerating the slicing process. Liu et al.~\cite{2021_1D-skeleton_Zouqiang_convolution-surface} proposed an out-of-core method to attain both memory- and time-efficient slicing of large-scale lattice structures. Noteworthily, the code of this method has been open-sourced. Fig.~\ref{fig:slicing_results}a shows some examples of the acceleration methods.
    
\textbf{Robustness Improvement.} To improve slicing robustness, existing methods consistently choose to use implicit representations, or called F-rep~\cite{2019_slicing_TPMS_CAD,2021_1D-skeleton_Zouqiang_convolution-surface,ding2021stl,2018_slicing_F-rep,2022_PersistentHomology,2023_slicing_persisitentHomology_adapative,2024_slicing_microcellular-structures,2024_slicing_Standard-Adaptive-cellular}. For example, Liu et al.~\cite{2021_1D-skeleton_Zouqiang_convolution-surface} introduced convolutional surfaces into microstructure modeling to improve slicing robustness. Dong et al.~\cite{2022_PersistentHomology,2023_slicing_persisitentHomology_adapative} introduced the persistent homology theory to guarantee the topological correctness of slices. Mustafa et al.~\cite{2021_slicing_efficiency_matrix-oriented-data-structure} provided a different way of improving slicing robustness, which converts the plane-lattice intersection problem into the easier line-ellipse intersection problem and then makes use of the existing ray tracing algorithm to carry out all the necessary calculations. This method is quite fast and easy to implement. Fig.~\ref{fig:slicing_results}b shows some examples of the robustness enhancement methods.

\subsubsection{Support generation}
\label{sec:support}
\begin{figure*}[t]
\centering
\includegraphics[width=0.94\textwidth]{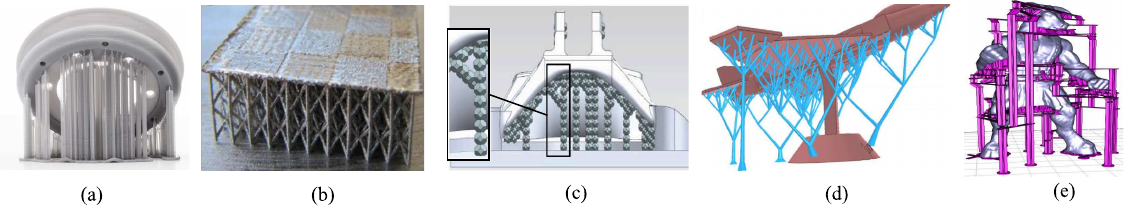}
\caption{Typical support structures: (a) column support; (b) lattice support~\cite{2013_support_Cloots_minimize-geometry_lattice-support}; (c) block-based support~\cite{2016_support_Vaidya_minimize-geometry_optimum-support}; (d) tree-like support~\cite{2019_support_Zhang_minimize-geometry_Local-Barycenter-Based}; and (e) bridge support~\cite{2014_support_Dumas_minimize-geometry_bridge-structure}.}
\label{fig:support_results_minimizing}
\end{figure*}

Support structures are used to prevent overhangs from dropping, sagging, curling, and even snapping off in a print. An overhang refers to any geometric portion that is unsupported from below, including horizontal/slanted protrusions, bridges between vertical walls, or internal cavities. For conventional solids, support structures are often generated \textit{a posteriori}, with two major steps. The first step is overhang identification through methods such as ray casting~\cite{2011_support_Qian_generic_STL-model-based} or slice difference~\cite{2013_slicing_LDNI}. Having found overhangs, the second step computes the support volume and fills it with appropriate structures, such as column support, zigzag support, gusset support, lattice support, and tree-like support~\cite{2013_slicing_intersection-free_topological-faithful,2014_support_Schmidt_minimize-geometry_branching-support,2014_support_Dumas_minimize-geometry_bridge-structure,2014_support_Vanek_geometry_Clever-Support,2019_support_Zhang_minimize-geometry_Local-Barycenter-Based,2019_support_Zhu_minimize-geometry_lightweight-tree-shaped,2019_support_Zhu_minimize-geometry_Tree-Shaped-Support,2019_support_Vaisser_minimize-geometry_Genetic-algorithm-based,2020_support_Zhang_minimize-geometry_Bio-inspired-tree-shape,2013_support_Hussein_minimize-geometry_lattice-support,2013_support_Cloots_minimize-geometry_lattice-support,2013_support_Strano_minimize-geometry_TPMS-support,2016_support_Gan_minimize-geometry_practical-support,2016_support_Vaidya_minimize-geometry_optimum-support}, as depicted in Fig.~\ref{fig:support_results_minimizing}. During generation, it is better to minimize support structure volumes for reduced material usage and printing time. Some commonly used methods for this purpose are:
\begin{itemize}
    \item \textbf{Printing orientation:} rotate the model so that overhangs are reduced or angled closer to the printing bed, as discussed in Sect.~\ref{sec:part_orienting}.    
    \item \textbf{Blending:} use blended edges instead of sharp ones to gradually change angles, which encourages self-supporting edges~\cite{2021_1D-skeleton_Zouqiang_convolution-surface}.
    \item \textbf{Model decomposition:} break the original model into multiple components with fewer overhangs and then assemble them together afterwards~\cite{2012_support_Luo_self-support-deformation_model-segmenting,2016_support_Wei_self-support-infill_partion}.
    \item \textbf{Topology optimization:} optimize support structures to reduce the support volume by topology optimization techniques~\cite{2019_support_Zhu_minimize-geometry_lightweight-tree-shaped,2019_support_Zhu_minimize-geometry_Tree-Shaped-Support,2019_support_Vaisser_minimize-geometry_Genetic-algorithm-based,2018_support_Mezzadri_minimize-support-optimization_mechanistic-meaning}. 
    \item \textbf{Empirical guidelines:} adjust protrusions to 45 degrees or less and bridges to 5mm or less in length whenever possible.
\end{itemize}

Unlike conventional solids, a microstructure's support structures not only can be generated \textit{a posteriori} but also \textit{a priori}. The latter way of working is possible because microstructures are usually designed using optimization techniques, and support structure design can be easily incorporated into this optimization-based framework as either additional objectives or constraints.

\begin{figure*}[t]
\centering
\includegraphics[width=0.8\textwidth]{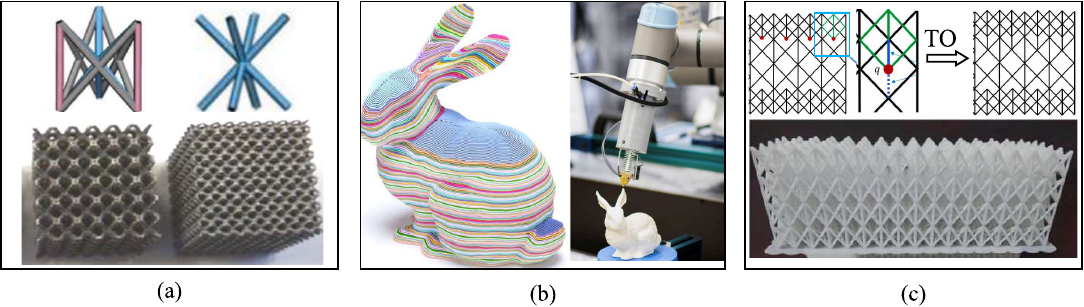}
\caption{Self-supported microstructures: (a) self-supported unit cells~\cite{2021_support_Zhou_self-support-design_multi-fold-rotational-symmetry}; (b) support-free multi-axis printing~\cite{2022_support_Zhang_self-support_multi-axis}; and (c) support-free 
 topology optimization~\cite{2023_support_Wang_self-support-deformation_subdivision-simplicification}.}
\label{fig:support_results_self-supported}
\end{figure*}

\textbf{A posterior methods.}  Basically, support generation methods of microstructures in this category follow the conventional methods outlined above. For example, Li et al.~\cite{2021_slicing_multi-directional} proposed to print lattice structures using multi-axis printing, which is essentially a continuous version of the printing orientation method. The model is constantly rotated to eliminate overhangs, and thus support structures. (It should be noted that, while multi-axis printing holds great promise, it also significantly increases the complexity of generating tool paths.) Also, all the typical support structures as shown in Fig.~\ref{fig:support_results_minimizing} have been applied to microstructures, e.g., column support for bio-inspired lattice structures~\cite{2022_support_column-for-lattice}, tree-like support for porous structures~\cite{2021_support_Wang_minimize-geometry_Support-point-determination} and lattice structures~\cite{2023_support_grid-tree-for-lattice}, etc. Another typical example is filling a part with microstructures respecting the empirical guidelines (the 45 degrees and 5mm rules), including Body-Centred Cubic unit cell~\cite{2016_support_Aremu_self-support-design_BCC}, TPMS unit cell~\cite{2019_support_Hu_self-support-design_TPMS}, and self-supported hierarchical unit cell~\cite{2021_support_Zhou_self-support-design_multi-fold-rotational-symmetry}, as depicted in Fig.~\ref{fig:support_results_self-supported}. However, this approach has the limitation of the restricted variety of available self-supported unit cells. Consequently, it becomes challenging to use this method for creating diverse microstructures that cater to various design requirements.

\textbf{A priori methods.} As their name suggests, methods in this category take support generation into consideration during the design stage, as objectives or constraints. (This is closely related to the notion of design for manufacturing.) These methods have been presented in various forms, but common to all is framing support generation as a topology optimization problem. For example, Wang et al.~\cite{2023_support_Wang_self-support-deformation_subdivision-simplicification} presented a topology optimization method to generate self-supporting lattice structures, based on two simple operations of subdivision and simplification. Similar ideas have also been presented in~\cite{2013_support_Wang_minimize_cost-effective,2017_support_Wang_self-support-infill_support-free-frame,2021_support_Liu_self-support-infill_TO-shell,2018_support_Lee_self-support-infill_ellipse-hollowing,2014_support_Leary_self-support-deformation_optimal-topology,2020_support_van_self-support-deformation_front-propagation,2018_support_Garaigordobil_self-support-deformation_new-overhang-constraint,2017_support_Langelaar_self-support-deformation_AM-filter,2016_support_Langelaar_self-support-deformation_3D-self-supporting}, to name a few. Noteworthy, Wu~\cite{2016_support_Wu_self-support-infill_rhombic} presented a rhombic structure-based topology optimization method that guarantees self-supportiveness.

Both approaches are currently employed, each with its advantages and challenges. A posteriori methods often rely on expert knowledge and manual intervention, lacking a principled way (i.e., still somewhat of an art) to generate support structures. By contrast, a priori methods are sound for the sake of using optimization techniques but face challenges in handling the large design space of complex microstructures.


\begin{figure*}[t]
\centering
\includegraphics[width=0.8\textwidth]{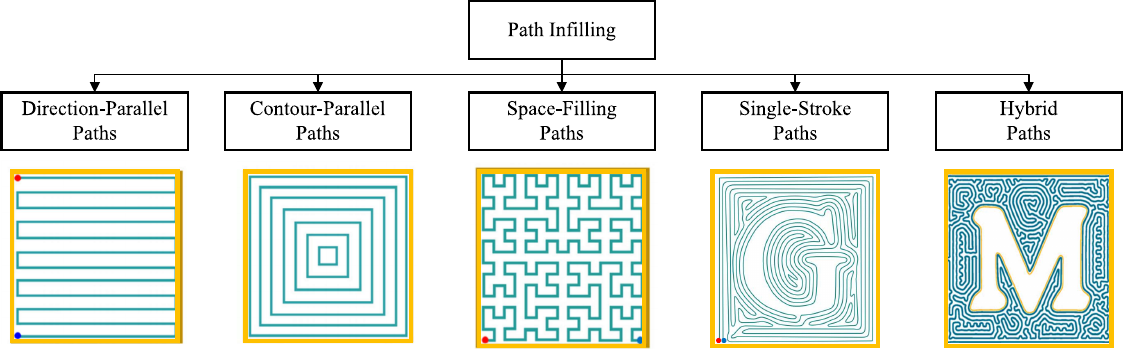}
\caption{Illustration of typical path patterns.}
\label{fig:path-pattern}
\end{figure*}

\subsubsection{Path infilling}
\label{sec:path_infilling}
Tool paths govern how a 3D printer moves its head relative to the printer bed during printing. Hence, the quality of tool paths directly impacts surface finish, speed, and strength of prints. From a purely geometric point of view, path infilling generates a series of paths within each slice so that printed material (characterized by printing strip widths) following those paths can completely cover the slice's interior region. In real practice, some physical factors are also important: the amount of material and time needs to be minimized, and enough strength and stability to withstand the printing process and the print's intended use also need to be ensured. These all mean that path infilling is an important and difficult task. Many methods have been proposed to tackle this problem, and they may be classified into five categories (same as those in conventional CNC machining): direction-parallel paths, contour-parallel paths, space-infilling paths, single-stroke paths, and their combinations. Fig.~\ref{fig:path-pattern} shows illustrations of these path patterns. For microstructures, the infilling paths basically follow these patterns, refer to Table \ref{tab:manufacturing-operation} for more details.

\textbf{Direction-parallel paths.} This strategy generates a series of (exactly or approximately) parallel paths along a specified direction, see the left of Fig.~\ref{fig:path-pattern}. The commonly used zigzag paths fall under this category. Because of their high computational efficiency, direction-parallel paths have been widely used (maybe the most used) in conventional solid printing/machining~\cite{zou2013iso,zou2021length}. Nonetheless, it should be noted that such paths are prone to sharp corners, which have a negative impact on printing speed and print strength. More importantly, the key consideration in generating direction-parallel paths lies in determining the optimal path direction~\cite{2000_Park_infill_direction-parallel,2001_Rajan_infill_zigzag,2015_Jin_infill_parallel-based-path}, but most methods treat this problem in an ad hoc manner.

\begin{figure*}[t]
\centering
\includegraphics[width=0.75\textwidth]{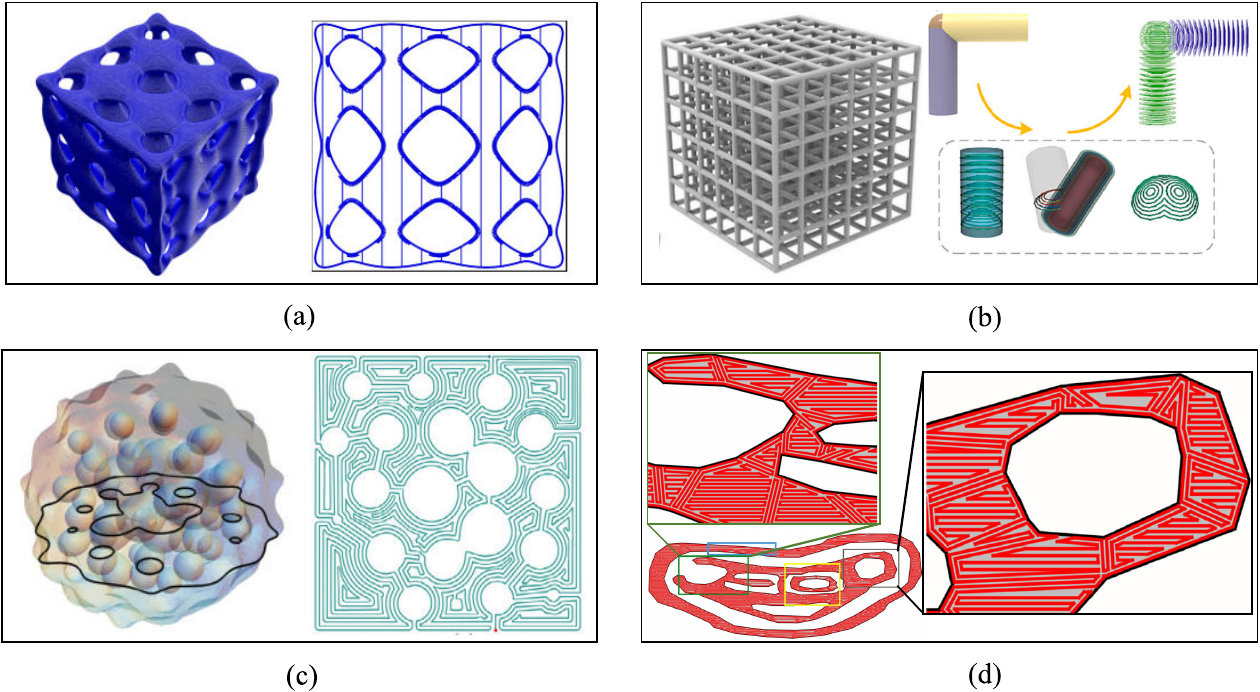}
\caption{Examples of infilling paths for microstructures: (a) directional-parallel paths~\cite{2018_slicing_F-rep}; (b) contour-parallel paths~\cite{2024_infill_truss-spherical-path-planning}; (c) single-stroke paths~\cite{2019_infill_Porous-Structures}; (d) contour-zigzag hybrid path~\cite{2022_hybrid-path_zigzag-contour}.}
\label{fig:path-results}
\end{figure*}

Direction-parallel paths have also found applications in microstructure printing. Ding et al.~\cite{ding2021stl} and Tang et al.~\cite{2024_direction-or-contour-path-TPMS} generated direction-parallel paths for shell-based microstructures. Huang et al.~\cite{2024_infill_direction-lattice} applied such path patterns for lattice structures. Also, direction-parallel paths have been applied to solid-based microstructures~\cite{2018_slicing_F-rep}, see Fig~\ref{fig:path-results}a. While direction-parallel paths maintain computational efficiency in microstructure applications, their limitations regarding the optimal path direction become more critical due to the intricacies of microstructures.

\textbf{Contour-parallel paths.} This strategy generates a series of paths parallel to a slice's contours (the boundaries between the solid and the void), see the middle left of Fig.~\ref{fig:path-pattern}. Such paths are usually calculated with the offset operation~\cite{rossignac1986offsetting,zou2014iso,zou2021robust}. In contrast to direction-parallel paths, contour-parallel paths are seen to have fewer sharp U-turns (and therefore less repeated acceleration and deceleration during printing)~\cite{2002_infill_Equidistant-path}. However, it is found that contour-parallel paths tend to produce small uncut regions and inflection points, resulting in under-fills and increased printing time.

In microstructure printing, contour-parallel paths are also widely used because of their smooth transitions, as shown in Fig~\ref{fig:path-results}b. Such path patterns have been employed for TPMS structures~\cite{2024_direction-or-contour-path-TPMS}, lattice structures~\cite{2024_infill_truss-spherical-path-planning}, and honeycomb-like structures~\cite{2019_infill_contour-complex-geometries,2023_infill_contour-continuous-honeycomb}. However, the challenge of small uncut regions and inflection points found in conventional prints becomes more pronounced for microstructure prints.

\textbf{Space-filling paths.} This strategy generates tool paths using space-filling curves, which, as the name suggests, are curves that can move through every point of space, see the middle of Fig.~\ref{fig:path-pattern}. Typical examples include spiral paths~\cite{2009_infill_space-filling_spiral}, Fermat paths~\cite{2016_infill_Fermat-Spirals}, and Hilbert paths~\cite{2017_infill_Nair_space-filling_Hilbert}. As can be seen from Fig.~\ref{fig:path-pattern}, space-filling paths (except for spiral paths) often have sharp corners, causing repeated acceleration/deceleration and under-fills.

Space-filling paths have also been applied to microstructures. For instance, Zhao et al.~\cite{2009_infill_space-filling_spiral} introduced a method for generating spiral paths for porous structures. While these paths effectively cover intricate shapes with high precision, the sharp corners associated with space-filling paths can pose significant challenges. Additionally, the complex contours of microstructure slices require more advanced algorithms to ensure that the paths accurately conform to these intricate contours.

\textbf{Single-stroke paths.}
Tool paths consisting of a single, continuous curve are sometimes preferred due to their high printing efficiency, see the middle of Fig.~\ref{fig:path-pattern}. These paths eliminate the need for frequent stops and starts, which reduces wear on the machine and improves both the surface quality and overall printing speed. A prime example is their application in the production of continuous carbon fiber reinforced thermoplastics~\cite{yamamoto2022novel}, where uninterrupted motion is critical for maintaining material consistency and structural integrity.

In microstructure printing, single-stroke paths have recently gained increasing attention~\cite{2022_contimuous_CGF,2019_infill_Porous-Structures,2023_infill_honeycome-infill,2020_infill_sparse-infill,2023_infill_contour-continuous-honeycomb}, refer to Fig~\ref{fig:path-results}c for examples. Because a slice of a microstructure consists of several components, the essential problem of generating single-stroke paths lies in continuous tool paths within individual components and short transitions among components. For the individual component tool paths, the aforementioned space-filling paths are often used~\cite{2016_infill_Fermat-Spirals}. (Note that, although space-filling paths are often presented as single-stroke paths, they can actually be multi-stroke paths with multiple starts and ends, and this is the reason why we distinguish between space-filling paths and single-stroke paths.) For inter-component transitions, the divide-and-conquer strategy~\cite{2019_infill_Porous-Structures} and Euler graphs have been used~\cite{2021_infill_lightweight-infill,2020_infill_sparse-infill,2023_infill_honeycome-infill}. Although the above methods are seen to generate effective tool paths, it is important to note that single-stroke paths tend to have complex, twisted local geometry, which may lead to heat accumulation (and consequently part distortion) and frequent changes in tool motion.


\textbf{Hybrid paths.} 
In certain scenarios, a single method may not suffice to generate high-quality tool paths since many factors are involved. This leads to the combined use of two or more generation methods, such as zigzag-contour paths~\cite{2011_zigzag-contour_conventional,2013_zigzag-contour_conventional}, which enhance both surface quality and infilling efficiency.

In microstructure printing, the zigzag-contour hybrid paths are frequently used~\cite{2022_Bi_infill_Continuous-contour-zigzag, 2022_Xia_infill_Continuous-contour-zigzag,2022_hybrid-path_zigzag-contour}, where slice contours are offset inward multiple times to create contour-parallel paths, and any remaining unfilled areas are covered by zigzag paths (Fig. 20d). This approach helps balance surface quality with printing time in microstructures. Besides this classical way of combining paths, Ozbolat et al.~\cite{2014_infill_hybrid_zigzag-spiral} introduced a new way to hybridize paths for porous structures, which follows a bilayer pattern of zigzag and spiral paths.

\begin{figure}[t]
\centering
\includegraphics[width=0.48\textwidth]{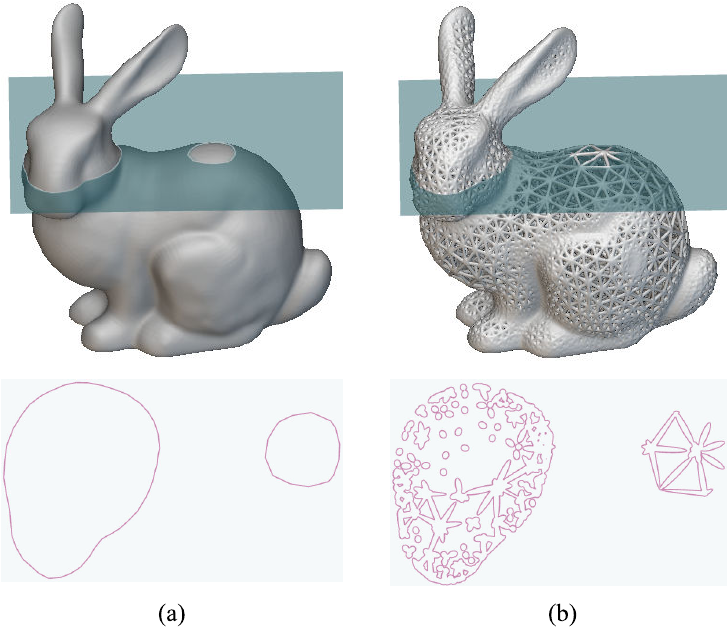}
\caption{Number of components of conventional solids (a) V.S. modern microstructures (b).}
\label{fig:layer_topology}
\end{figure}

As can be seen from the discussions above, tool path generation methods for microstructures are almost no different from those for conventional solids. Most methods are just direct applications of conventional methods. One notable exception is that a microstructure's slice usually has hundreds of thousands of components while a conventional solid's slice has only several components, see Fig.~\ref{fig:layer_topology}; therefore, generating single-stroke paths is much more difficult for microstructures than for conventional solids, and current component transition methods~\cite{2022_contimuous_CGF,2019_infill_Porous-Structures,2023_infill_honeycome-infill,2020_infill_sparse-infill} are insufficient to handle this challenge.

\section{Summary on progress, limitations and opportunities}
\label{sec:summary}
The previous two sections have compiled research studies related to microstructure modeling. In Sect.~\ref{sec:challenges}, there are at least four major challenges identified for microstructure modeling: representational compactness, computational efficiency, computational robustness, and multiscale integrity. Here, a summary of existing methods' limitations, and therefore future research opportunities, regarding the four challenges is to be outlined. 

\subsection{Representational compactness}
\label{sec:summary-represention} 
\textbf{What has been done.}
The representation of microstructures can be categorized into topological and geometric representations. Topological representation concerns the arrangement of cells within a microstructure, while geometric representation focuses on the specific shape of these cells. Progress in topological representation includes:
\begin{itemize}
    \item For regular topology, which admits a periodic arrangement pattern, program-based methods have been developed (see Sect.~\ref{sec:regular}). These methods compactly store regular microstructures using programs of several KB, capable of representing million-scale or even billion-scale microstructures.
    \item For semi-regular topology, which has a local periodicity or follows a predefined variation pattern, many effective methods have been developed (see Sect.~\ref{sec:semi-regular}), and common to them is storing how the topology developed/varied, not the final topology directly (e.g., a predefined variation pattern).
    \item For irregular topology, which has stochastical arrangements, current methods are still in their early stage of development (see Sect.~\ref{sec:irregular}). The topology is simply stored as graph structures, which cannot handle large-scale microstructures.
\end{itemize}

Geometric representation has made significant advancements over topological representation. Both explicit representations and implicit representations, as well as their hybrids, have been researched for a long time (see Sects.~\ref{sec:1D}-\ref{sec:4D}). Progress in this direction is summarized as follows:
\begin{itemize}
    \item For 1D beam-based cell geometries, curve-based methods are used, and the curves can be parametric, explicit, and implicit (see Sect.~\ref{sec:1D}). They are, however, limited to linear or quadratic curves for the sake of easy computation.
    \item For 2D shell-based cell geometries, surface-based methods have been developed, and the surfaces can be parametric, explicit, and implicit (see Sects.~\ref{sec:2D} and~\ref{sec:4D}). If explicit surfaces are to be used, the geometry is often kept simple (linear or quadratic) for easy computation; if using implicit surfaces, the geometry can be quite complex.
    \item For 3D solid-based cell geometries, many effective methods have been developed, including voxels, texture mapping, volumetric splines, and scalar fields, see Sects.~\ref{sec:3D} and~\ref{sec:4D}. These methods can handle complex cell geometries.
\end{itemize}

\textbf{What remains open.} 
Despite the significant research progress stated above, several serious problems remain unsolved. Topology representation methods are compact and general for regular microstructures but lack general solutions for semi-regular microstructures and effective solutions for irregular microstructures, particularly large-scale ones (CS and HCS). Geometric representation schemes, while capable of handling small-scale microstructures, suffer from high memory consumption when scaling up to cases of CS/HCS.

\textbf{What to expect.}
It is safe to say that current representation schemes are adequate for regular or small-scale microstructures but will become problematic if semi-regular/irregular and large-scale microstructures are considered. (This may explain why demonstrations of most existing research studies on microstructure modeling and design are all about small-scale microstructures, many even on toy examples.) To overcome this limitation, there are several directions to explore, among which the following two are noteworthy:
\begin{itemize}
    \item Compressive representation. Reducing redundant information (e.g., repeating topological and geometric patterns) in microstructure models is an effective way to improve representational compactness. To this end, 3D compression can play an important role in microstructure representation. The primary challenges lie in (1) an effective metric to measure topological and geometric complexity (which allows similarity comparisons), (2) a fast recognition algorithm to identify similar topological and geometric patterns in a large-scale microstructure, and (3) effective encoding and decoding of similar topological and geometric patterns.
    \item On-demand generative representation. To further improve representational compactness, we may go from storing low-level topological/geometric details to storing high-level generation semantics. Similar to biology, which uses genomes to encode how the geometry is generated, not the final geometry, we can embed representation into a set of generation algorithms that are efficient to store. These generation algorithms must work locally so that topological/geometric details can be generated in an on-demand manner. That is, the user/calling algorithm specifies initial conditions like locations and functions to drive the generation algorithms to return intended local topological/geometric details. These details can be stored, visualized, and processed using current methods since they are local and at a small scale. For instance, Gupta et al.~\cite{2019_3D-rep_Gupta_CSG_steady-lattice}, Kurzeja and Rossignac~\cite{2022_Rossignac_Tran-Similar}, and Rossignac~\cite{2020_procedural_COTS} have provided a typical strategy for such on-demand generation, where specific portions of lattice structures are generated by applying transformations to base unit cells, rather than evaluating the entire microstructure.
\end{itemize}

\subsection{Computational efficiency}

\textbf{What has been done.}
The microstructure modeling algorithms may be divided into design-oriented operations and manufacturing-oriented operations. In short, design-oriented operations refer to those used to create, analyze, and edit a microstructure model, while manufacturing-oriented operations are for process planning. Their respective research progress is summarized as follows:
\begin{itemize}
    \item For design-oriented operations, only a small portion of conventional CAD/CAM operations have been implemented for microstructures, including querying, boundary evaluation, blending, offsetting, simplifying, conforming, booleans, and hollowing, see Sect.~\ref{sec:design-operation}. Among them, blending, offsetting, and conforming operations are maturer than the others, a consequence of many readily applicable methods from previous CAD/CAM research, e.g., F-rep modeling and volumetric splines. 
    \item For manufacturing-oriented operations, better research progress than design-oriented operations has been seen. All the important steps (part orienting, slicing, support generation, and path infilling) in the microstructure manufacturing pipeline have efficient algorithms made available, see Sect.~\ref{sec:manufacturing-operation}. There are good reasons for this progress: (1) most of the manufacturing-oriented operations are just customized versions of general-purpose ones used in the conventional CAD/CAM systems; and (2) process planning is an offline task, meaning that efficiency is not a big issue here~\footnote{This is not to imply that efficiency is not important to manufacturing-oriented operations, but that no high efficiency (e.g., realtime) is required.}.
\end{itemize}

\textbf{What remains open.}
Most notably, many design operations, which have proven useful for CAD/CAM design process, have not been implemented for microstructures. One typical example is direct modeling~\cite{zou2019push,zou2022robust-direct,2019_3D-rep_Gupta_CSG_steady-lattice}. This direct, graphical approach to manipulating microstructure models can significantly enhance intuitiveness and efficiency in microstructure modeling, as demonstrated by Gupta et al.~\cite{2019_3D-rep_Gupta_CSG_steady-lattice}. Even for the operations already adapted to microstructure modeling, some are still far from being practically useful, e.g., querying, boundary evaluation, simplifying, and booleans. Fast querying operations have only been developed for some special kinds of microstructures, which cannot even be applied to general regular microstructures, let alone semi-regular and irregular ones. Boundary evaluation operations are still too slow and not robust for large-scale microstructures. Consequently, it is still impossible for current CAD/CAM systems to interactively visualize large-scale microstructures. Another example goes to Boolean operations, whose current implementations are too compute-intensive to be useful in SS and HCS modeling.

\textbf{What to expect.}
Like the situation in microstructure representation, the most urgent and challenging problem of microstructure computation is also handling large-scale microstructures, especially semi-regular and irregular microstructures. After all, small-scale microstructures can at least be handled as conventional solids using existing CAD/CAM algorithms. To increase the efficiency of large-scale microstructure computation, GPU parallel computing techniques may be helpful. Modern GPUs are well-suited for efficiently processing large-scale microstructures due to the massive parallelism they offer. However, applying GPUs to microstructures is no trivial matter---simply allocating one GPU thread per computation task does not necessarily mean those threads will execute in parallel due to memory and warp divergence~\footnote{The detailed reason for this is modern GPUs use warps (typically consisting of 32 threads) as operation primitives. All threads in a warp share the same instructions and memory access. For this reason, if we want all threads in a warp to execute simultaneously, coalesced memory access and balanced workloads must be explicitly designed, otherwise, those threads will execute serially (one waiting for another to finish)~\cite{2020_Wang_memory-divergence}.}, two problems rarely studied in the CAD/CAM community. To this end, two major challenges exist~\cite{zou2024meta}: (1) a new GPU-friendly data structure and a coalesced-access memory layout for storing microstructures in GPUs with minimized memory divergence; and (2) workload-balanced scheduling algorithms to reduce warp divergence. In addition, data transfer between CPUs and GPUs is in general slow, and thus a compressed microstructure representation is also needed to reduce data transfer overheads.

\subsection{Computational robustness}
\textbf{What has been done.}
As previously noted, many microstructure modeling algorithms are adaptations of general-purpose algorithms from the traditional CAD/CAM domain. These adaptations primarily focus on addressing computational efficiency challenges, while relatively little attention has been given to robustness issues. The limited research on modeling robustness often adopts a straightforward strategy: directly applying existing implicit modeling techniques to enhance robustness in manipulating microstructure models, as demonstrated in studies such as~\cite{2021_1D-skeleton_Zouqiang_convolution-surface}.

\textbf{What remains open.}
For microstructures, degeneracies---such as coincident or self-intersected geometries---are more common than in conventional solid models, posing significant challenges for robust computation. These issues can often be more complex than those encountered in traditional CAD/CAM modeling. For instance, Boolean and blending operations, which are already challenging in conventional solid modeling, become even more problematic when applied to microstructures. From the authors' experience using Siemens NX and CATIA to model microstructures (in particular TPMS models), over 90\% of Boolean and blending operations failed. While other software packages, such as nTopology, offer improved performance, they are limited to implicit modeling and face interoperability issues with the standard CAD/CAM pipeline based on B-rep models~\cite{zou2024meta}. Despite the critical importance of the robustness issues, they have received relatively little attention.

\textbf{What to expect.}
Generally, implicit modeling is robust but memory- and compute-intensive. Explicit modeling is more efficient but often has robustness issues, e.g., blending and booleans. Hybridizing them is seemingly a promising way to attain high computational robustness without sacrificing efficiency. Little research effort has, however, been devoted toward this direction. The primary challenges lie in (1) an automatic and efficient conversion mechanism between the two modeling schemes to keep representational consistency, and (2) a memory-efficient data structure to minimize the overheads caused by storing multiple representations. Note that it is only feasible to store multiple representations of a microstructure's local geometry, rather than the overall geometry, a consequence of the model data explosion problem even for a single representation. The on-demand generative representation method discussed earlier may help mitigate this issue.

\subsection{Multiscale integrity}
\label{sec:Multiscale-integrity}
\textbf{What has been done.}
Unlike preceding topics, maintaining integrity between geometric details at different scales in microstructure models remains largely unexplored. To date, no dedicated methods have been publicly documented. The closest relevant research involves the conforming operations discussed in Sect.~\ref{sec:design-operation}. This operation allows micro-level cell geometries and meso-level topology of a microstructure to be conformed to its macro-level shape. However, they primarily address initial microstructure generation but do not extend to situations of microstructure editing.

\textbf{What to expect.}
Achieving multiscale integrity poses a significant challenge, particularly in managing interactions between geometric changes and physical properties. That is, when propagating the geometric changes from one scale to the other, we need to ensure that physical properties are respected during the changes. For instance, when modifying a macrostructure's boundary shape, it is essential to update micro-scale cell geometries to prevent artifacts like dangling or isolated pieces (Fig.~\ref{fig:consistency}), and meanwhile maintaining prescribed physical property, e.g., density fields.
The development of methodologies that seamlessly propagate geometric modifications across scales while respecting physical constraints is a critical area for future research. Reverse design from physical properties to geometric specifications plays a pivotal role, and artificial intelligence techniques may help in this regard.

\section{Conclusion}
\label{sec:conclusion}
This review has provided a comprehensive and state-of-the-art overview of geometric modeling methods for microstructure design and manufacturing. It aims to clarify the challenges and progress of current research and provides references for subsequent research. The review began with an identification of four major challenges in microstructure modeling: representational compactness, computational efficiency, computational robustness, and multiscale integrity (Sect.~\ref{sec:challenges}). 
Then, a systematic classification of microstructures is provided (Sect.~\ref{sec:taxonomy}), according to their geometric and topological characterizations at the macro-, meso-, and micro-level. 
Based on these challenges and classification, a thorough review of published studies on microstructure representations and operations was conducted (Sects.~\ref{sec:representation} and~\ref{sec:operation}). 
The strengths and limitations of these methods were evaluated in the context of the identified challenges, highlighting significant research progress in microstructure modeling. A notable limitation identified is the current difficulty in modeling large-scale microstructures (Sect.~\ref{sec:summary}). 
The review concludes with some promising directions that may be expected in the near future, particularly the compressive and generative microstructure representation schemes, as well as parallel and hybrid microstructure modeling algorithms.

It is important to acknowledge that, despite our efforts to comprehensively cover existing methods in this review, the extensive literature on this topic may have resulted in unintentional omissions of relevant contributions. We apologize for any oversights and encourage further exploration to advance the field of geometric modeling methods for microstructure design and manufacturing.

\section*{Acknowledgements}
This work has been funded by NSF of China (No. 62102355), the ``Pioneer" and ``Leading Goose" R\&D Program of Zhejiang Province (No. 2024C01103), NSF of Zhejiang Province (No. LQ22F020012), and the Fundamental Research Funds for the Zhejiang Provincial Universities (No. 2023QZJH32).


\section*{References}

\bibliographystyle{elsarticle-num}
\bibliography{bibfile}




\end{document}